\documentclass[aps,prl,twocolumn,showpacs,10pt,preprintnumbers,nofootinbib]{revtex4-2}

\usepackage{amsmath, amssymb, amsfonts, mathtools, mathrsfs, slashed,xfrac}
\numberwithin{equation}{section}

\usepackage{array, makecell, multirow, booktabs, hhline, colortbl}

\usepackage{graphicx, rotating, epstopdf, epsfig}
\usepackage[dvipsnames]{xcolor}

\usepackage[compatibility=false]{subcaption}
\usepackage{latexsym, MnSymbol, braket}

\usepackage[utf8]{inputenc}
\usepackage{orcidlink}
\usepackage{contour}
\usepackage{comment}

\usepackage{tikz-feynman}
\usetikzlibrary{
    shapes,
    shapes.geometric,
    shapes.symbols,
    shapes.arrows,
    shapes.multipart,
    shapes.callouts,
    shapes.misc,
    decorations.pathmorphing,
    arrows.meta}
\tikzset{
    snake it/.style={decorate, decoration=snake},
    cross/.style={cross out, draw=black, minimum size=2*(#1-\pgflinewidth), inner sep=0pt, outer sep=0pt},
    cross/.default={1pt}
}
\tikzfeynmanset{warn luatex=false}
\usetikzlibrary{decorations.markings}

\usepackage{hyperref}
\hypersetup{
    colorlinks=true,
    linkcolor=black,
    citecolor=black,
    filecolor=black,
    urlcolor=black,
    pdfborder={0 0 0}
}

\usepackage{eso-pic}

\newcommand{\En}[1]{\mathcal{E}({#1})}

\newcommand{\LO}{\mathbb{O}}
\newcommand{\n}{\mathbf{n}}
\newcommand{\p}{\mathbf{p}}
\newcommand{\LIR}{\Lambda_{\text{IR}}}

\newcommand{\xn}{x_{\mathbf{n}}}
\newcommand{\xnp}{x_{\mathbf{n}}^+}
\newcommand{\xnm}{x_{\mathbf{n}}^-}
\newcommand{\xnT}{x_{\mathbf{n}}^T}
\newcommand{\xnd}{\hat{x}[\n]}
\newcommand{\xnpd}{\hat{x}_{\mathbf{n}}^+}
\newcommand{\xnmd}{\hat{x}_{\mathbf{n}}^-}
\newcommand{\BK}[1]{\mathcal{K}_{#1}}

\begin{document}

\title{Energy Correlators in Warped Geometries
}

\author{Lorenzo Ricci~\orcidlink{0000-0001-8704-3545}}
\affiliation{Maryland Center for Fundamental Physics, Department of Physics,
	University of Maryland, College Park, MD 20742, USA}
\author{Raman Sundrum~\orcidlink{0009-0004-7537-5357}}
\affiliation{Maryland Center for Fundamental Physics, Department of Physics,
	University of Maryland, College Park, MD 20742, USA}

\begin{abstract}
\noindent  
We study Energy Correlators as probes of strongly-coupled nearly-conformal field theories within their holographically dual descriptions, focusing on the important features that appear in realistic theories going beyond the standard model. In particular, 
we study warped geometries which asymptote to $\text{AdS}_5$, as well as IR-truncations dual to a 4D gap. Our correlators are computed by in-in type Witten perturbative diagrams, corresponding to a large-$N$ expansion of the strong dynamics. We describe how this sets the stage for phenomenological applications for collider searches beyond the standard model as well as for new theoretical explorations in Lorentzian holography. 
\end{abstract}

\maketitle

\tableofcontents
\section{Introduction}
It is highly plausible that  physics beyond the standard model (BSM) includes strongly-coupled dynamics, which can generate the hierarchical structure of particle physics we observe. For example, Higgs compositeness and partial compositeness of other standard model particles are elegant strong-coupling mechanisms of this type (reviewed in \cite{Gherghetta:2010cj}). It is a major goal of collider experiments to discover such BSM physics. 
Of course,  strong coupling presents a theoretical challenge to understand its non-perturbative features and a phenomenological challenge to associate measurements with theoretically meaningful quantities. 
QCD provides an interesting  and subtle precedent, greatly simplified by asymptotic freedom at collider energies while still producing highly non-perturbative hadronic features at detector scales. However, asymptotic freedom is not guaranteed beyond the standard model and it is a robust possibility that BSM physics is strongly coupled even at collider energies, most straightforwardly if it is close to a non-trivial RG fixed point in the UV so the dynamics is approximately conformal there. 
This conformal regime can however be theoretically tractable if it has  a weakly-coupled AdS/CFT holographic dual. 
The weak couplings in the higher-dimensional AdS description are dual to a large-$N$ type expansion of the CFT.
The AdS theory is characterized by the mass-spin spectrum of AdS$_5$ particles and their interactions, roughly dual to CFT scaling dimensions and OPE coefficients. It would be very exciting to discover such BSM physics and measure these fundamental AdS/CFT data. Energy correlators are the ideal observables for pragmatically formulating this task phenomenologically. 

Energy Correlators (ECs) were first proposed in the context of QCD measurements at the LEP collider \cite{Basham:1979gh, PhysRevD.17.2298}. The basic idea is to measure energy deposited in calorimeters as a function of angle around a collision point, independent of other details of the specific final state. 
This defines a natural set of IR/Collinear-safe observables. In QCD, these are precisely calculable in asymptotically free perturbation theory. Intuitively, decreasing angles probe physics at decreasing energy scales in units of the hard scale in the underlying process, so that confinement effects are important at sufficiently small angles, where perturbative calculability is therefore lost.
Recently, there has been a revival of interest for applications to LHC physics.   More theoretically, energy correlators have been developed in the context of (strongly coupled) CFT, starting from the seminal paper of Ref.~\cite{Hofman:2008ar}. See Ref.~\cite{Moult:2025nhu} for a comprehensive review of these developments and an extensive list of references.

However, applications to strongly-coupled BSM physics remain largely unexplored. In this paper, we begin developing some of the theoretical tools to pursue this goal. The issues we tackle are of interest theoretically to better understand {\it Lorentzian} AdS/CFT duality, as well as being relevant for phenomenology. In a realistic setting, the strong coupling physics will not be exactly conformal, but will in general have a non-trivial RG flow. Furthermore, the IR of the strong dynamics will necessarily have a mass gap. This is obvious if the strong sector is a UV extension above the standard model (SM), such as in composite Higgs dynamics, since the weakly-coupled SM must emerge by currently observed energies. But even if the strong dynamics appears in a ``hidden sector'' without SM charges it is crucial to have a gap below which hidden states decay back to the SM so that they can be detected by SM calorimeters. Finally, the strong dynamics will not be isolated, but weakly coupled to other fields which necessarily deform away from conformality. The holographic dual of all these aspects is captured by warped compactifications, which take the form of Randall-Sundrum I (RS1) models \cite{Randall:1999ee} in higher-dimensional EFT, consisting of an approximately AdS$_5$ bulk truncated by UV and IR boundaries.  

The presence of a gap means that the strong dynamics (RS1) has a well-defined 4D S-matrix unlike exact CFT. Yet, this S-matrix is hard to calculate. At strong coupling a small gap relative to collider energy will typically result in a high-multiplicity final state from cascade decays which cannot be captured at fixed order even within a weakly-coupled RS1 dual. Furthermore, in AdS (RS1 bulk) EFT the heavy AdS fields that have been integrated out are dual to composite high-dimension CFT operators which interpolate 4D states which can still be light enough (in 4D mass) to be produced on-shell. Yet these states are not captured by the AdS effective description. Energy correlators provide  partially inclusive  physical observables which can mitigate these problems, recovering holographic perturbative EFT control.  

Most of the existing work on holographic energy-correlators exploits the magic of conformal symmetry,  which is broken in realistic settings.  Here, we begin the perturbative framing of these holographic correlators in more general warped geometries in which only 4D Poincare symmetry is exact. Energy detector operators are constructed from the energy-momentum tensor of the strong dynamics, dual to 5D General Relativity in the holographic description. When the strong dynamics is a CFT the detectors can effectively be placed at null infinity and are then dual to AdS shockwave geometries which then leads to  dramatic simplifications \cite{Hofman:2008ar}. 
However, these simplifications are lost or obscured once we break exact conformal invariance. In particular,  in the presence of a 4D mass gap final state particles cannot formally reach null infinity. 
Furthermore, non-trivial angular correlations between different detectors requires bulk interactions of the incoming state beyond those minimally required with the detector gravitational field, as we will show is true even beyond the conformally symmetric limit. 
These extra interactions do not reduce to simple plane-wave (free particle) propagation in the shockwave geometry of the detectors. 

Because of these limitations when going beyond conformal invariance 
we study EC directly in terms of in-in variants of Witten diagrams in the holographic description, working in 4D momentum space and in position space in the holographic direction. We find this to better clarify the physics and kinematics of EC in collider settings and connects to usual 4D perturbative calculations at weak coupling.\footnote{ This sacrifices some of the advantages of combining a completely position space approach with conformal symmetry \cite{Belitsky:2013xxa}.} We will comment on how the Witten diagrams capture simplifications of the shockwave approach in the gapless limit. 
In future work, we wish to apply this framework to collider phenomenology, either searching for hidden sectors in the ongoing LHC  (and its high-luminosity upgrade) or searching for  UV extensions of the standard model at future colliders.

Refs.~\cite{Csaki:2024joe,Csaki:2024zig,Csaki:2025abk} have studied holographic energy correlators motivated by 
 ``AdS/QCD'' models of QCD and the confinement gap in particular. While these employ similar frameworks to RS1, here we are inspired by BSM realizations and we also use a quite different methodology.
 Presently, we find significant disagreement on several of the overlapping issues. We will comment on these in Section~\ref{Sec:RS}. 

Our paper is organized as follows. In Section~\ref{Sec:AdS}, we start in exact AdS and derive ECs for a simple one-particle AdS state, dual to a single-trace CFT state in the large-$N$ expansion. 
At leading order in large-$N$ this particle only interacts with the detectors. Our results match with known results, but with a rather simple physical picture. However, more general particle interactions (dual to $1/N$ corrections) are needed in order to get non-trivial angular correlations between different detectors. We study a simple example in Section~\ref{Sec:Inter}, exploiting the angular operator product expansion, showing how one can measure AdS masses (CFT scaling dimensions). In Section~\ref{Sec:IrrDef}, we study the deformation away from exact AdS (dual to non-trivial renormalization group (RG)  flows), while in Section~\ref{Sec:RS} we study the impact of an IR boundary as in RS1 (dual to a 4D mass gap). We present our outlook in Section~\ref{Sec:Outlook}. 

We end this introduction by briefly defining the basic operator observable that will be relevant for the paper and to present our notation.
 
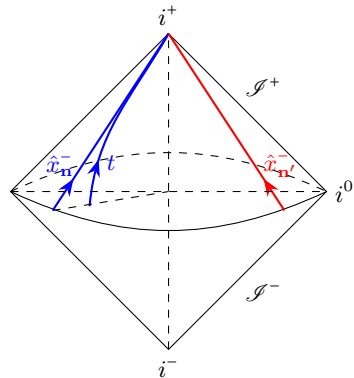
\begin{figure}
    \centering
    \begin{tikzpicture}[scale=.7]

        \draw (-3,0) -- (0,3) -- (3,0) -- (0,-3) -- cycle;

        \draw (-3,0) to[out=-25,in=-155] (3,0);
        \draw[dashed] (-3,0) to[out=25,in=155] (3,0);

        \draw[thick,blue,
  postaction={decorate},
  decoration={markings, mark=at position 0.25 with {\arrow{Stealth}}}
] (-1.5,-0.28) to[out=85,in=-120]
  node[pos=0.22, right] {$t$}
  (0,3);

\draw[thick,blue,
  postaction={decorate},
  decoration={markings, mark=at position 0.18 with {\arrow{Stealth}}}
] (-2.2,-0.36) --
  node[pos=0.26, left] {$\xnmd$}
  (0,3);
  \draw[thick,red,
  postaction={decorate},
  decoration={markings, mark=at position 0.18 with {\arrow{Stealth}}}
] (2.2,-0.36) --
  node[pos=0.26, right] {$\hat{x}_{\n'}^-$}
  (0,3);
        \draw[dashed] (-2.2,-0.36) -- (0,0);
        \draw[dashed] (-3,0) to (3,0);
        \draw[dashed] (0,-3) to (0,3);
        

        \node[above] at (1.8,1.65) {$\mathscr{I}^+$};
        \node[below] at (1.8,-1.55) {$\mathscr{I}^-$};

        \node[above] at (0,3) {$i^+$};
        \node[below] at (0,-3) {$i^-$};
        \node[right] at (3,0) {$i^0$};

    \end{tikzpicture}
    \caption{Penrose representation of Mikowski space, showing the two different contour of integration for detectors. The curved blue line represents the worldline of integration of a detector at finite distance from the central collision point (see \eqref{Eq:EnDetector}). The red and blue straight lines represent the light-ray limit of detectors worldline applicable for gapless 4D theories (see \eqref{Eq:EnDetectorLC}).}
    \label{fig:DetInt}
\end{figure}

\textit{\textbf{Energy Correlators}}
Energy detectors are modeled as operators,  
\begin{align}\label{Eq:EnDetector}
    \En{\n}=\lim_{r\rightarrow + \infty} r^{2} \int_0^{+\infty} dt \, T^{0 i}(t, r \n)\, n_{i}\,.
\end{align}
which measure the energy flux in the direction $\n$ at a large distance $r\rightarrow \infty$ from the ``collision point''.
In gapless theories such as CFTs, where signals propagate asymptotically at the speed of light, the integration contour can equivalently be pushed to null infinity $\mathscr{I}^+$ (see, e.g., \cite{Belitsky:2013xxa}), giving the equivalent form
\begin{align}\label{Eq:EnDetectorLC}
    \En{\n}=\frac{1}{2^4}\lim_{\xnpd\rightarrow + \infty} (\xnpd)^{2} \int_{-\infty}^{+\infty} d\xnmd \,
    T^{\mu \nu}(\xnd)\,\bar{n}_{\mu }\bar{n}_{\nu}\,,
\end{align}
where we defined the vector $\xnd = \frac{\xnpd}{2} n^{\mu}+\frac{\xnmd}{2} \bar{n}^{\mu}$ with $n^{\mu} = (1,\n)$ and $\bar{n}^{\mu}=(1,-\n)$. Fig.~\ref{fig:DetInt} illustrates the two integration contours which approach each other for large $r\rightarrow\infty$.

For momentum energy with positive energy, $\ket{s(p)}$,\footnote{More precisely, the momentum eigenstates appearing in \eqref{Eq:GenEEC} should be understood as normalizable wave packets of the form
\begin{equation}
\int d^4x \, e^{-i p \cdot x}\,
e^{-\frac{(x^0)^2+\mathbf{x}^2}{2 \sigma^2}}\,
s(x)\,|0\rangle ,
\end{equation}
with the limit $\sigma \to \infty$ taken at the end of the calculation. We will comment on this point whenever relevant in the rest of the paper.
} we define \emph{Energy Correlators} as Energy detector ``in-in'' expectations,
\begin{align}\label{Eq:GenEEC}
    \braket{\En{\n_1}\ldots \En{\n_N}}_{s(p)}
    = \frac{ \braket{s(p)|\En{\n_1}\ldots \En{\n_N}|s(p)}}{\braket{s(p)|s(p)}}\,.
\end{align}
The definition \eqref{Eq:GenEEC} further relies on the fact that energy detectors commute with the momentum generator,
$[P^{\mu},\En{\n}]=0$ as we will verify for our cases of interest.\footnote{This can be shown by a direct calculation in ordinary weakly coupled field theory, see e.g.~\cite{Bauer:2008dt}. In a CFT, instead, the relations $[P^{\mu}, \En{\n}]=0$ follow from the fact that $\En{\n}$ transforms as a primary operator inserted at spatial infinity~\cite{Kravchuk:2018htv}.} 

\textbf{Notation}. 
We use capital letters $M,N,\ldots$ for 5D indices, with 
$X^{M}=(x^{\mu},w)$. We adopt mostly minus convention. Given a unit three–vector $\mathbf{n}$ representing a direction in ordinary three-space, we introduce the null vectors
\begin{align}
n^{\mu}=(1,\mathbf{n}), \qquad \bar n^{\mu}=(1,-\mathbf{n}),
\end{align}
and define the light–cone decomposition for any for four-vector $v^{\mu}$
\begin{align}\label{Eq:PMNotation}
    v^{\mu} 
    = \frac{v_{\n}^+}{2}\, n^{\mu}
    + \frac{v_{\n}^-}{2}\, \bar n^{\mu}
    +(v_{\n}^T)^{\,\mu},
\end{align}
where the transverse component satisfies 
$n\cdot v_{\n}^T = \bar{n}\cdot v_{\n}^T = 0$. Detector positions are a special case notationally.
We denote them, prior to the limit in \eqref{Eq:EnDetectorLC}, by $\hat{x}^{\mu}[\n]$, with components
\begin{align}
    \hat{x}^{\mu}[\n]
    =
    \frac{\hat{x}_{\n}^+[\n]}{2}\, n^{\mu}
    +
    \frac{\hat{x}_{\n}^-[\n]}{2}\, \bar{n}^{\mu}\,,
\end{align}
while $x_T^{\mu}[\n] = 0$. With a slight abuse of notation, we will drop the $[\n]$ argument in the components $\hat{x}[\n]_{\n}^{\pm}$ and simply denote them by $\hat{x}_{\n}^{\pm}$.

\begin{figure*}[t]
\centering

\newcommand{\biggap}{1.2em}
\newcommand{\smallgap}{0.4em}

\begin{minipage}[t]{0.22\textwidth}
\centering
\begin{tikzpicture}[scale=0.7,baseline=(current bounding box.center)]
  \draw (-2,-1.5) -- (-2,2);
  \draw (-1,-2) -- (-1,1.5);
  \draw (-2,-1.5) -- (-1,-2);
  \draw (-2,2) -- (-1,1.5);
  \draw[-{Stealth}] (-1,-2) -- (2.5,-2);
  \node at (2.5,-2.4) {$w$};
  \node at (-1.7,1.1) {$\mathcal{E}$};
  \begin{feynman}
    \vertex at (0.85,0) (b1);
    \vertex at (1.6,-0.25) {$w=\frac{p\,\xnp}{p_{\n}^+}$};
    \vertex at (-0.8,-1) (a2);
    \vertex at (-1.5,1) (c1);
    \vertex at (1.5,1.1) (c2);
    \diagram*{ (b1) --[photon] (c1), };
  \end{feynman}
  \draw[thick,
    postaction={decorate},
    decoration={markings, mark=at position 0.3 with {\arrow{Stealth}}}
  ] (a2) to[out=20,in=-120,looseness=1.2] (c2);
\end{tikzpicture}
\captionof{figure}{Holographic one-point EC for a classical Point-Particle in AdS.}
\label{fig:PP_One}
\end{minipage}
\hspace{\biggap}
\begin{minipage}[t]{0.74\textwidth}
\centering

\begin{minipage}[c]{0.31\textwidth}
\centering
\begin{tikzpicture}[scale=0.7,baseline=(current bounding box.center)]
  \draw (-2,-1.5) -- (-2,2);
  \draw (-1,-2) -- (-1,1.5);
  \draw (-2,-1.5) -- (-1,-2);
  \draw (-2,2) -- (-1,1.5);
  \draw[-{Stealth}] (-1,-2) -- (2.5,-2);
  \node at (2.5,-2.4) {$w$};
  \node at (-1.4,1.3) {$\mathcal{E}$};
  \node at (-1.8,0.8) {$\mathcal{E}$};
  \begin{feynman}
    \vertex at (1,0.25) (b1);
    \vertex at (0.5,-.3) (b2);
    \vertex at (-0.8,-1) (a2);
    \vertex at (-1.3,1.1) (c1);
    \vertex at (-1.7,0.5) (c3);
    \vertex at (1.5,1.1) (c2);
    \diagram*{ (b1) --[photon] (c1), (b2) --[photon] (c3)};
  \end{feynman}
  \draw[thick,
    postaction={decorate},
    decoration={markings, mark=at position 0.3 with {\arrow{Stealth}}}
  ] (a2) to[out=20,in=-120,looseness=1.2] (c2);
\end{tikzpicture}

\vspace{0.3em}
{\small (a)}
\end{minipage}
\hspace{\smallgap}
\begin{minipage}[c]{0.31\textwidth}
\centering
\begin{tikzpicture}[scale=0.7,baseline=(current bounding box.center)]
  \draw (-2,-1.5) -- (-2,2);
  \draw (-1,-2) -- (-1,1.5);
  \draw (-2,-1.5) -- (-1,-2);
  \draw (-2,2) -- (-1,1.5);
  \draw[-{Stealth}] (-1,-2) -- (2.5,-2);
  \node at (2.5,-2.4) {$w$};
  \node at (-1.4,1.3) {$\mathcal{E}$};
  \node at (-1.8,0.8) {$\mathcal{E}$};
  \begin{feynman}
    \vertex at (0.85,0) (b1);
    \vertex at (-0.8,-1) (a2);
    \vertex at (-1.3,1.1) (c1);
    \vertex at (-1.7,0.5) (c3);
    \vertex at (1.5,1.1) (c2);
    \diagram*{ (b1) --[photon] (c1), (b1) --[photon] (c3)};
  \end{feynman}
  \draw[thick,
    postaction={decorate},
    decoration={markings, mark=at position 0.3 with {\arrow{Stealth}}}
  ] (a2) to[out=20,in=-120,looseness=1.2] (c2);
\end{tikzpicture}

\vspace{0.3em}
{\small (b)}
\end{minipage}
\hspace{\smallgap}
\begin{minipage}[c]{0.31\textwidth}
\centering
\begin{tikzpicture}[scale=0.7,baseline=(current bounding box.center)]
  \draw (-2,-1.5) -- (-2,2);
  \draw (-1,-2) -- (-1,1.5);
  \draw (-2,-1.5) -- (-1,-2);
  \draw (-2,2) -- (-1,1.5);
  \draw[-{Stealth}] (-1,-2) -- (2.5,-2);
  \node at (2.5,-2.4) {$w$};
  \node at (-1.4,1.3) {$\mathcal{E}$};
  \node at (-1.8,0.8) {$\mathcal{E}$};
  \begin{feynman}
    \vertex at (0.85,0) (b2);
    \vertex at (0,0.3) (b1);
    \vertex at (0.08,0.55){$Y$};
    \vertex at (-0.8,-1) (a2);
    \vertex at (-1.3,1.1) (c1);
    \vertex at (-1.7,0.5) (c3);
    \vertex at (1.5,1.1) (c2);
    \diagram*{ (b1) --[photon] (c1), (b1) --[photon] (c3),(b1)--[photon](b2)};
  \end{feynman}
  \draw[thick,
    postaction={decorate},
    decoration={markings, mark=at position 0.3 with {\arrow{Stealth}}}
  ] (a2) to[out=20,in=-120,looseness=1.2] (c2);
\end{tikzpicture}

\vspace{0.3em}
{\small (c)}
\end{minipage}

\captionof{figure}{Holographic two-point EC for a classical Point-Particle in AdS. We show that the two contributions in panels (b) and (c) vanish.}
\label{fig:PP_Two}
\end{minipage}

\end{figure*}

\section{Detectors and holography in $\text{AdS}$}\label{Sec:AdS}
We begin by considering ECs applied to a CFT which enjoys a weakly-coupled $\text{AdS}_5$ dual, with background metric
\begin{equation}
ds^2 \equiv \bar{G}_{MN}dX^{M}dX^N =\frac{dx^2 - dw^2}{w^2}\,.
\end{equation}
We work in units where the AdS radius is set to unity, $R_{\text{AdS}} \equiv 1$.
 We consider an AdS effective field theory (EFT) consisting of 
 a scalar field $\Phi$ coupled to 5D GR, 
\begin{align}\label{Eq:AdsScalar}
  S=\int d^4 x\, dw \sqrt{G}
  \left(
  \frac{1}{2}\partial_{M} \Phi\,\partial^{M} \Phi
  - \frac{M^2_{\Phi}}{2} \Phi^2
  +\frac{ R - 2 \Lambda}{2 \kappa}
  \right)\,.
\end{align}

We will take the state $\ket{s(p)}$ in \eqref{Eq:GenEEC} of a single $\Phi$ propagating in AdS. In Sec.~\ref{Sec:AdSPP} we will treat this 5D-massive particle in the classical Point-Particle (PP) limit, deferring  a second quantized treatment of $\Phi$ until Sec.~\ref{Sec:AdSSQ}. The important physics will be more transparent in this way.

\subsection{Lightning derivation: Classical Point Particle}\label{Sec:AdSPP}

The AdS Point-Particle is described by the action
\begin{align}\label{Eq:PPAction}
    \mathcal{S}_{\text{PP}}
    = \frac{1}{2} \int d \tau
    \left(
        \frac{1}{e}\, \dot{X}^{M} G_{MN} \dot{X}^{N}
        + e\, M_{\Phi}^2
    \right)\,,
\end{align}
where $e\equiv e(\tau)$ is the worldline einbein and dots denote derivatives with respect to the worldline parameter~$\tau$.

Varying the action \eqref{Eq:PPAction} with respect to $e$ yields the constraint
\begin{align}\label{Eq:PPConstraint}
    G_{MN}\dot{X}^M \dot{X}^N - e^{2} M_{\Phi}^2 = 0\,.
\end{align}
Using reparameterization invariance, we can freely fix the gauge $e M_{\Phi} = 1$. Together with the equations of motion for $X^M$, this leads to
\begin{align}\label{Eq:PPmomenta}
    p_{\mu} = M_{\Phi} \frac{\dot{x}_{\mu}}{w^2}\,,
    \qquad
    \dot{w}^2 = \frac{w^2}{M_{\Phi}^2}\left(p^2 w^2 - M_{\Phi}^2\right)\,,
\end{align}
where $p_\mu$ is the conserved four-momentum along the usual 4D directions.

Notice that the consistency of the second equation in \eqref{Eq:PPmomenta} requires
\begin{equation}
    p\, w > M_{\Phi}\,,
\end{equation}
implying that smaller values of the holographic coordinate $w$ are not classically accessible.

It is convenient to parametrize the trajectory as a function of $w$. Using
\begin{align}
    \frac{dx^{\mu}}{dw}
    = \frac{\dot{x}^{\mu}}{\dot{w}}
    = \frac{p^{\mu} w}{\sqrt{p^2 w^2 - M_{\Phi}^2}}\,,
\end{align}
we obtain
\begin{align}\label{Eq:PPtraj}
    x^{\mu}_{\text{PP}}(w)
    = \frac{p^{\mu}}{p^2}
      \sqrt{p^2 w^2 - M_{\Phi}^2}\,,
\end{align}
where we fixed the integration constant by imposing the boundary condition
$x^{\mu}_{\text{PP}}(w = M_{\Phi}/p) = 0$, and we defined $p \equiv \sqrt{p^2} > 0$.

We are first interested in the expectation value of a single energy detector,
\begin{align}\label{Eq:DetectoPPLC}
\braket{\En{\n}}_{\text{PP}}
&=
\frac{1}{2^4}
\lim_{\xnpd \rightarrow \infty}
(\xnpd)^2
\int_{- \infty}^{\infty} d\xnmd \,
\bar{n}_{\mu}\bar{n}_{\nu}
\braket{T^{\mu \nu} (\xnd)}_{\text{PP}}\,,
\end{align}
as in \eqref{Eq:DetectoPPLC} where $\braket{T^{\mu \nu} (x)}_{\text{PP}}$ is given holographically as
\begin{align}\label{Eq:TmunuHolo}
    \braket{T^{\mu \nu}(x)}_{\text{PP}}
    =
    \int \frac{d^4 y\, dw}{w^5}\,
    K^{\mu \nu}_{\alpha \beta}(x-y,w)\,
    \mathcal{T}_{\text{PP}}^{\alpha \beta}(y,w)\,.
\end{align}
Here $\mathcal{T}_{\text{PP}}^{MN}(x,w)$ is the stress tensor of the point-particle
\begin{align}
    \mathcal{T}^{MN}_{\text{PP}}(x,w)
    &=
    \frac{2}{\sqrt{G}}
    \frac{\delta \mathcal{S}_{\text{PP}}}{\delta G_{MN}(x,w)} \\
    &=
    \int d \tau \,
    \frac{1}{\sqrt{G}}\,
    \dot{X}^{M}(\tau)\dot{X}^{N}(\tau)\,
    \delta^{(5)}\!\left(X - X(\tau)\right)\,,\notag
\end{align}
and $K^{\mu \nu}_{\alpha \beta}(x,w)$ is the bulk-to-boundary propagator in AdS for the metric
perturbation that we conveniently express in mixed position-momentum space
\begin{align}\label{Eq:GravBB}
    K_{\mu \nu,\alpha\beta}(q,w)
    =
\frac{q^2}{2}\BK{2}(qw)\Pi_{\mu \nu,\alpha \beta}(q)
\end{align}
with $\Pi_{\mu \nu, \alpha \beta}$ the projector onto the symmetric
traceless tensor components, $\BK{2}$ the Bessel function of the second kind and where we are working in axial gauge: $G_{5M}-\bar{G}_{5M}=0$. Notice that to obtain the correct retarded (classical) prescription \eqref{Eq:GravBB} is actually evaluated at $q=\sqrt{\mathbf{q}^2 - (q^0+ i \epsilon)^2}$ and that the normalization of \eqref{Eq:TmunuHolo} with \eqref{Eq:GravBB} is such that the external stress-tensor measures energy.

Combining \eqref{Eq:DetectoPPLC} and \eqref{Eq:TmunuHolo}, we can perform the
$\xnm$ integral first. 

Referring back to \eqref{Eq:PMNotation} for the notation, we find
\begin{align}\label{Eq:LCIntgral}
    &\int d\xnmd \int \frac{d^4 q}{(2 \pi)^4} \,
    e^{i q\cdot (\xnd - x_{\text{PP}})} \frac{q^2}{2}\BK{2}(qw)\Pi^{++}_{\,\,\,\alpha \beta}(q)=\notag\\
    &\delta(\xnpd - x_{\text{PP},\n}^+)\!\!\underbrace{\int\!\! \frac{d^2 q_{\n}^T}{(2 \pi)^2}\,
    e^{i q_{\n}^T\cdot x_{\text{PP},\n}^T}
    \frac{(q_{\n}^T)^2}{2}\BK{2}(w\sqrt{(q_{\n}^T)^2})}_{\sim\frac{w^2}{\left((\xnT)^2+w^2\right)^3}}\delta^+_{\alpha}\delta^+_{\beta}\,.
\end{align}
The equality follows from the fact that the $\xnmd$ integral enforces
$q_{\n}^{+}=0$. As a consequence, any dependence on $q_{\n}^{-}$ in the
integrand, including that appearing in the polarization tensor $\Pi^{++}_{\,\,\alpha \beta} (q)$, drops out, except for the phase factor.
The subsequent integration over $q_{\n}^{-}$ then yields the delta
function $\delta(\xnpd - x_{\text{PP},\n}^{+})$. Notice that this result
relies crucially on the fact that the $\xnm$ integration is performed
along a lightlike direction.

From the classical trajectory \eqref{Eq:PPtraj}, we see that the large-$\xnp$
limit corresponds to large values of the holographic coordinate $w$, where the
five-dimensional mass $M_{\Phi}$ can be neglected. In this regime, the trajectory
simplifies to $ x_{\text{PP}}^{\mu}(w) \simeq w \,p^{\mu}/p$. In the same limit, the point-particle stress tensor reduces to
\begin{align}\label{Eq:StressTensorMassles}
    \mathcal{T}_{\text{PP}}^{\mu \nu}(x,w)
    \xrightarrow[w\to\infty]{}
    \frac{w^5}{w^2 p}\,
    w^4 p^{\mu} p^{\nu}\,
    \delta^{(4)}\!\left(
    x^{\alpha} - \frac{p^{\alpha} w}{p}
    \right)\,.
\end{align}

The remainder of the calculation is now straightforward. After combining \eqref{Eq:DetectoPPLC} with \eqref{Eq:TmunuHolo} and performing the light-ray integration as in \eqref{Eq:LCIntgral} the integral over the interaction point is completely saturated by the $\delta$-function 
\begin{align}\label{Eq:SingDetPP}
    \braket{\En{\n}}_{\text{PP}} =\frac{1}{4 \pi} \int \frac{dw}{pw^2}  \delta( \xnpd-x_{\text{PP},\n}^+) \frac{(\xnp)^2w^6 (p_{\n}^+)^2}{((x_{\text{PP},\n}^T)^2+w^2)^3} \,,
\end{align}
and we finally obtain 
\begin{align}\label{Eq:EPPresult}
    \braket{\En{\n}}_{\text{PP}}
    =
    \frac{1}{4\pi}\,
    \frac{(p^2)^2}{(n\!\cdot\! p)^3}\,.
\end{align}

The result in \eqref{Eq:EPPresult} agrees with the direct calculation of the ECs in a CFT. While our state $\ket{s(p)}$ is described by a single Point-Particle in AdS, it holographically corresponds to a strongly coupled CFT state with center-of-mass (CoM) momentum $p^{\mu}$ which is spreading over time as capture by $w\rightarrow\infty$ as $t\rightarrow \infty$. Because the AdS Point-Particle represents a spread out state from the CFT perspective, notice that $\p$ has not to be pointed in the same direction of $\n$ in order to be registered by the detector. In other words the CoM does not to have to head directly to the detector in order for some part of the CFT state to be detected.

Consider now two detectors, $\En{\n_1}$ and $\En{\n_2}$. The calculation consists
of three contributions, depicted in Fig.~\ref{fig:PP_Two}. The first one is
trivially factorized, 
\begin{align}
\braket{\En{\n_1}\En{\n_2}}_{\text{PP}}^{(2a)}
&=
\braket{\En{\n_1}}_{\text{PP}}\,
\braket{\En{\n_2}}_{\text{PP}}\,,
\end{align}
because the point-particle trajectory is fixed by \eqref{Eq:PPtraj} and each detector is then given by two copy of \eqref{Eq:TmunuHolo} with independent integrals over the PP trajectory in \eqref{Eq:PPtraj}.

However the remaining two
contributions, shown in Fig.~\ref{fig:PP_Two}~(b,c), could in principle lead to a
non trivial correlation between detectors and so non trivial correction to II.17. Yet, in AdS these connected contributions are known to vanish, as first
pointed out in Ref.~\cite{Hofman:2008ar}. We will explain this point later, in
Sec.~\ref{Sec:IrrDef}, and further generalize it to the case of asymptotic AdS geometries.

Furthermore, this result straightforwardly generalizes to an arbitrary number of
detectors,
\begin{align}\label{Eq:FactoPPFull}
    \braket{\En{\n_1}\ldots \En{\n_N}}_{\text{PP}}
    &=
    \braket{\En{\n_1}}_{\text{PP}}\cdots
    \braket{\En{\n_N}}_{\text{PP}}\\
    &=\frac{1}{(4 \pi)^N} \frac{(p^2)^2}{(n_1
\cdot p)^3}\cdots  \frac{(p^2)^2}{(n_N
\cdot p)^3}\,.\notag
\end{align}

This factorization is a consequence of the tree-level approximation in AdS, which is dual to a large-N color limit from the CFT view point. This result will not survive loop corrections in AdS dual to $1/N$ corrections in the CFT. These loop corrections will further introduce new non-trivial dependencies on angular separations between detectors parametrized by $n_i\cdot n_j$, even for fixed $n_i \cdot p$. In other words, in the rest frame $p=(p^0, \mathbf{0})$, \eqref{Eq:FactoPPFull} does not depend on the angles between the detectors but this property will be spoiled by loops.

\subsection{Second quantization}\label{Sec:AdSSQ}

We now consider the same problem in the second-quantized description, which is more useful when considering general interactions. 

It is convenient to decompose the bulk scalar field as (see Ref.~\cite{Sundrum:2011ic} for a thorough discussion)
\begin{align}\label{Eq:ContKKDecomposition}
    \Phi(x,w)
    =
    \int_{0}^{\infty} \!\! dm \, \mathcal{N}_m
    \int d\Omega_{\mathbf{p}}^{m}\,
    f_m(w)
    \left(
        a_{m}^{\dagger}(\mathbf{p})\, e^{i p\cdot x}
        + \text{h.c.}
    \right),
\end{align}
where
\begin{align}
    f_{m}(w) = w^2 \, J_{\nu}(m w)\,,
    \qquad
    d\Omega_{\mathbf{p}}^{m}
    \equiv
    \left.
    \frac{d^3 \mathbf{p}}{(2\pi)^3\, 2\omega_{\mathbf{p}}}
    \right|_{\omega_{\mathbf{p}}=\sqrt{\mathbf{p}^2+m^2}}\,,
\end{align} 
$\mathcal{N}_{m} = \sqrt{m}$ and $\nu=\sqrt{M_{\Phi}^2+4}$.
The mode function $\mathcal{N}_mf_{m}(w)$ satisfy the completeness relation
\begin{align}\label{Eq:ModeNorm}
    \int_0^{\infty} dm \, m \, f_{m}(w)\, f_{m}(w')
    =
    w^3\delta(w-w')\,,
\end{align}
while the annihilation and creation operators $a^{(\dagger)}_m(\mathbf{p})$
obey the usual canonical commutation relations with continuum  normalization,
\begin{align}\label{Eq:SecQuant}
    [a_{m}(\mathbf{p}),a_{m'}^{\dagger}(\mathbf{p}')]
    &=
    (2\pi)^3 2\omega_{\mathbf{p}}\,
    \delta^{(3)}(\mathbf{p}-\mathbf{p}')\,
    \delta(m-m')\,, \notag\\
    [a_{m}(\mathbf{p}),a_{m'}(\mathbf{p}')]
    &=
    [a_{m}^{\dagger}(\mathbf{p}),a_{m'}^{\dagger}(\mathbf{p}')]
    = 0\,.
\end{align}

We now consider a state with positive energy $p^0>0$
\begin{align}
\begin{aligned}
    \ket{\Phi(p,w)} &\equiv \int d^4x e^{-i p\cdot x} \Phi(x,w)\ket{0} =\pi \sqrt{p}f_{p}(w) a_p^{\dagger}(\mathbf{p})\ket{0}\,,
\end{aligned}    
\end{align}
localized in the $w$ direction. Notice that for $w>0$ this is not a diffeo-invariant state but it is consistent to consider it in the presence of diffeo gauge-fixing, as typically done in pertubative calculations. It reduces to a diffeo-invariant state as $w\rightarrow0$. But here we start with $w>0$ because this will be a useful intermediate building block for higher order calculations. Furthermore we have
\begin{align}
    \int d^4x e^{ipx}&\braket{0|\Phi(x,w_f)\Phi(0,w_i)|0}\equiv \braket{\Phi(p,w_f)|\Phi(p,w_i)}\notag\\
    &\qquad\qquad=\theta(p^0) \theta(p^2) \pi f_p(w_f)f_{p}(w_i)\,.
\end{align}

Let us focus first on the one point EC
\begin{align}\label{Eq:OnePointEC}
\braket{\Phi(p,w_{f})|\En{\n}|\Phi(p,w_i)}\,,
\end{align}
where, now, the stress-tensor at the boundary corresponds to the following bulk operator
\begin{align}\label{Eq:LocalStressTensor}
    \mathcal{E}(\n) =\frac{1}{2^4}\!\lim_{\xnpd\rightarrow\infty}&(\xnpd)^2 \int\!\!d\xnmd \int \!\!\frac{dwd^4y}{w^5}\times \\ &\times\bar{n}_{\mu}\bar{n}_{\nu}K^{\mu\nu}_{\,\,\alpha\beta}(\xnd-y,w)\mathcal{T}_{\Phi}^{\alpha \beta}(y,w)\,, \notag
\end{align}
with $\mathcal{T}_{\Phi}^{MN}$ the 5D stress tensor for $\Phi$, expressed in terms of second quantized fields.

Practically \eqref{Eq:OnePointEC} can be computed starting from the following ``in-in'' Witten-diagram 
\begin{align}\label{Eq:OnePointDiagram}
\begin{aligned}
    \begin{tikzpicture}[scale=0.7]
\draw (0, 0) circle (2cm);
\node at (0,2.4) {$\bar{n}^{\mu}\bar{n}^{\nu}T_{\mu\nu}(q)$};
\begin{feynman}
\vertex at (0,0) (b1);
\vertex at (-1.5,0) (a1);
\vertex at (-1.5,-0.3) {$w_i$};
\vertex at (1.5,-0.3) {$w_f$};
\vertex at (0,-0.3) {$w$};
\vertex at (1.5,0) (a2);
\vertex at (0,2) (c1);
\diagram*{
(a1) -- [momentum = $p$] (b1), 
(b1) -- (a2),
(c1) -- [photon, momentum= $q$] (b1),
}; 
\end{feynman}
\end{tikzpicture}
\end{aligned} =\begin{aligned}\pi f_{p}(w_i)f_{p+q}(w_f) I(p,q,w)\end{aligned}\,.
\end{align}
where
\begin{align}\label{Eq:1PointFull}
    I(p,q,w) = \pi f_{p}(w) f_{p+q}(w) \bar{n}^{\mu}\bar{n}^{\nu}K_{\mu \nu,\alpha \beta}(q,w) V^{\alpha \beta}(p,q)\,.
\end{align}
and $V_{MN}(p,q)$ is the graviton-scalar-scalar vertex. Notice that at this stage, before performing the light-ray integral and sending the detector to infinity, there is a finite momentum $q$ flowing in the graviton line. Furthermore, we stress that in (\ref{Eq:OnePointDiagram},~\ref{Eq:1PointFull}) the bulk-to-bulk propagator for the external state(s) are effectively \textit{on-shell}, meaning that $f_p$ denotes $f_m$ with $m = \sqrt{p^2}$. This is a consequence of the in-in structure of the correlator. We also emphasize that the circle in \eqref{Eq:OnePointDiagram} does not indicate global AdS; it simply denotes that points inside the circle correspond to bulk fields, while points on the circle represent boundary fields, according to the usual AdS/CFT prescription.

We are interested in \eqref{Eq:OnePointDiagram} where the detector is inserted at the position $\xnd$ (via Fourier transform) and we perform the operations on $\xnd$ as in \eqref{Eq:EnDetectorLC},
\begin{align}\label{Eq:1PointInt} \lim_{\xnpd \rightarrow\infty}\!\!(\xnpd)^2 \!\!\!\int \!\!d \xnmd \!\!\!\int\!\!\!\frac{d^4q}{(2 \pi)^4}\!\!\!\int \!\!\frac{dw}{w^5}e^{iq\cdot\xnd} f_{p+q}(w_f)I(p,q,w)\,.
\end{align}
The reader can check that this expression has a finite large-$\xnp$ limit, where we scale $1/q^{\mu} \sim w \sim \xnp$. Then, the mode function in \eqref{Eq:1PointFull} scales as 
\begin{align}
\label{Eq:AdsPlaneWave}
f_p(w)\, \mathop{\longrightarrow}\limits_{w\to\infty}
w^2 \sqrt{\frac{2}{\pi}} \frac{\cos(p w)}{\sqrt{p w}}
\sim (\xn^+)^{5/2}\, .
\end{align}
Meanwhile, the bulk-to-boundary graviton propagator scales as
$K_{\mu\nu,\alpha \beta}(q,w) \sim 1/(\xnp)^2$,\footnote{This can be readily understood by starting in position space, where $K(\lambda x,\lambda w)\sim \lambda^{-6} K(x,w)$, and then stripping out an overall $\lambda^{-4}$ through the Fourier transform.}
while the vertex scales as $V^{\alpha\beta}(p,q)\sim (\xnp)^4$ (noting the raised indices).  
It then follows that $I(p,q,w) \sim (\xnp)^5$, while the full expression in \eqref{Eq:1PointInt} is finite once $\xnm$ also scales linearly with $\xnp$. Finally, note that the external mode function $f_{p+q}(w_f)$ simply loses its $q$ dependence in this limit and does not scale with $\xnp$.

We thus get, for large $\xnp$,
\begin{align}
   \int \frac{dw}{w^5} \int\frac{d^4q}{(2 \pi)^4}e^{iq\cdot\xn} I(p,q,w)\simeq \notag\\
    \int_0^{\infty} \frac{dw}{p w^2}\int \frac{d^4q}{(2 \pi)^4} e^{i q \cdot(\xn^{\mu}-p^{\mu} \frac{w}{p})}\bar{n}^{\mu}\bar{n}^{\nu} K_{\mu \nu,\alpha \beta} (q,w)V^{\alpha\beta}(p,0)\,,
\end{align}
and we can immediately recognize that the Fourier transform of $K_{\mu \nu,\alpha \beta}$ is precisely the expectation value of the stress tensor on the classical trajectory we found in \eqref{Eq:LCIntgral}. \footnote{The $\cos(pw)$ in \eqref{Eq:AdsPlaneWave} comes with positive and negative phases. We kept only the combination which does not leave a large phase in the $\xnpd \rightarrow \infty $ limit.} The rest of the calculation parallels the classical Point Particle and we get
\begin{align}\label{Eq:IIntegralSol}
  \lim_{\xnpd\rightarrow\infty}(\xnpd)^2 \!\!\!\int d\xnmd\!\!\int\!\! \frac{dw}{w^5} \!\!\!\int\!\!\frac{d^4q}{(2 \pi)^4}e^{iq\cdot\xnd} I(p,q,w)=\frac{(p^2)^2}{(p\cdot n)^3}\,,
\end{align}
and we finally get
\begin{align}\label{Eq:1PointBulk}
    \frac{\braket{\Phi(p,w_f)|\En{\n}|\Phi(p,w_i)}}{\braket{\Phi(p,w_f)|\Phi(p,w_i)}}
    &=  \frac{1}{4 \pi}\frac{(p^2)^2}{(n \cdot p)^3}\,.
\end{align}

\begin{figure}
\centering
\begin{tikzpicture}[scale=0.7]
\draw (0, 0) circle (2cm);
\node at (0,2.3) {$\En{\n}$};
\begin{feynman}
\vertex at (0,0) (b1);
\vertex at (-1.5,0) (a1);
\vertex at (1.5,0) (a2);
\vertex at (0,2) (c1);
\diagram*{
(a1) -- [momentum = $p$] (b1) -- [] (a2),
(b1) -- [photon] (c1),
}; 
\end{feynman}
\end{tikzpicture}
\begin{tikzpicture}[scale=0.7]
\draw (0, 0) circle (2cm);
\begin{feynman}
\vertex at (0,0) (b1);
\vertex at (-1.5,0) (a1);
\vertex at (1.5,0) (a2);
\vertex at (0,2) (c1);
\diagram*{
(a1) -- [momentum = $p$] (b1) --  (a2)
}; 
\end{feynman}
\draw[dashed] (0,-1.5) -- (0,1.5);
\node at (3.2,0){$\times \frac{(p^2)^2}{(4 \pi)(n\cdot p)^3}$};
\node at (-2.4,0){$=$};
\end{tikzpicture}
\caption{Cutting relation for the one-point EC discussed in the text. Here, $n^{\mu} =(1 , \n)$.}
    \label{fig:WittenCut}
\end{figure}
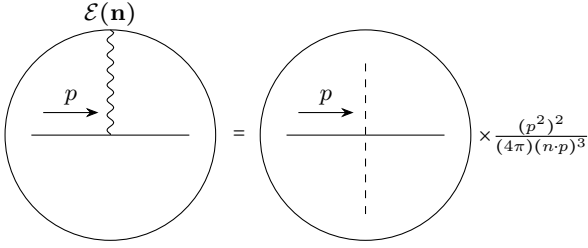

It is instructive to notice that, diagrammatically, we have derived the Feynman rule shown in
Fig.~\ref{fig:WittenCut}, which closely resembles the usual structure encountered
in 4D flat-space weakly-coupled theories and its relationship to the cutting prescription. Indeed, once the bulk propagator is written in mixed
momentum–position space $(p,w)$, the cutting rules are completely analogous to
the standard ones (see, e.g., Ref.~\cite{Meltzer:2020qbr}
for a thorough discussion). The Feynman propagator reads
\begin{align}\label{Eq:Bulktobulkpropagator}
    G_F(p,w_f,w_i)
    =
    i \int_0^\infty dm \, m \,
    \frac{f_m(w_i)\, f_m(w_f)}{p^2 - m^2 + i \epsilon}\,.
\end{align}
The cutting rule acts on the denominator $(p^2 - m^2 + i \epsilon)$ according to
the usual Minkowski Cutkosky prescription. Notice that, in the present case, the denominator in
\eqref{Eq:1PointBulk} is itself the cut propagator, which cancels against the
corresponding numerator.

We now move on and consider two detectors
\begin{align}\label{Eq:TwoPointEEC}
    \braket{\Phi(p,w_f)|\En{\n_1}\En{\n_2}|\Phi(p,w_i)}\,.
\end{align}
Following the previous discussion we have just to consider the following ``InIn'' diagram
\begin{align}\label{Eq:2PointDiagram}
\begin{aligned}
    \begin{tikzpicture}[scale=0.7]
\draw (0, 0) circle (2cm);
\node at (-1.5,2.3) {$\bar{n}_{1}^{\mu}\bar{n}_{1}^{\nu}T_{\mu \nu}(q_1)$};
\node at (1.8,2.3) {$\bar{n}_{2}^{\mu}\bar{n}_{2}^{\nu}T_{\mu \nu}(q_2)$};
\begin{feynman}
\vertex at (-0.5,0) (b1);
\vertex at (0.5,0) (b2);
\vertex at (-1.5,0) (a1);
\vertex at (-1.5,0.3) {$w_i$};
\vertex at (1.5,-0.3) {$w_f$};
\vertex at (-0.2,0.3) {$w_1$};
\vertex at (0.5,-0.3) {$w_2$};
\vertex at (1.5,0) (a2);
\vertex at (-1.41,1.41) (c1);
\vertex at (1.41,1.41) (c2);
\diagram*{
(a1) -- [momentum'=$p$] (b1), 
(b1) -- [] (a2),
(c1) -- [photon] (b1),
(c2) -- [photon] (b2),
}; 
\end{feynman}
\end{tikzpicture}
\end{aligned} \!\!\!\!\!\!\!\!= \!\begin{aligned} \pi f_{p}(w_i)f_{p+q_1+q_2}(w_f) \times\\I(p,q_1,w_1)I(p+q_1,q_2,w_2)\end{aligned}\,,
\end{align}
where $I(p,q,w)$ was defined before in \eqref{Eq:1PointFull}. Notice again that all the scalar lines are on-shell, in the sense that the mass index of the mode function matches the invariant mass of the corresponding state, since their creation and annihilation operators are simplified in terms of (product of) their commutators. 

The calculation of \eqref{Eq:TwoPointEEC} via \eqref{Eq:2PointDiagram} follows closely the previous discussion. We can start by performing the integrals in the region $1/q_{1}^{\mu} \sim w\sim \xnp\rightarrow\infty$ keeping fixed $q_2$ and $w_2$. After the integral we get $I(p,q_1) = (p^2)^2/(n_1\cdot p)^3$ as in \eqref{Eq:IIntegralSol}. Then we can perform the second integral which ultimately results in 
\begin{align}\label{Eq:TwoPointECCRes}
    \frac{\braket{\Phi(p,w_f)|\En{\n_1}\En{\n_2}|\Phi(p,w_i)}}{\braket{\Phi(p,w_f)|\Phi(p,w_i)}} = \frac{1}{(4 \pi)^2}\frac{(p^2)^2}{(n_1\cdot p)^3}\frac{(p^2)^2}{(n_2\cdot p)^3}\,.
\end{align}

In principle there are two other diagrams accompanying \eqref{Eq:2PointDiagram}, which are the second-quantized analogs of Figs.~\ref{fig:PP_Two}(b) and (c), but as stated in Sec.~\ref{Sec:AdSPP} these diagrams vanish, as explained in Sec.~\ref{Sec:IrrDef}. Therefore \eqref{Eq:TwoPointECCRes} is the complete answer for two detectors.

This result straightforwardly generalizes to several detectors and we can conclude that the action of the detector is just
\begin{align}\label{Eq:AdSEnOp}
    \En{\n} = \int_
    {0}^{\infty} dm\int d\Omega_{\mathbf{p}}\,a^{\dagger}_m (\mathbf{p}) a_m (\mathbf{p}) \frac{1}{4 \pi}\frac{p^4}{(n\cdot p)^3}\,,
\end{align}
i.e. just \emph{weighting} free-particles in the bulk by the single-particle response.

Notice also that localizing the states on the boundary via the usual AdS/CFT prescription, i.e. considering the state created by the CFT operator $\mathcal{O}$ dual to $\Phi$
\begin{align}\label{Eq:TwoPointEEC}
    \ket{\mathcal{O}(p)} \equiv \lim_{w\rightarrow0} \frac{1}{w^{\Delta_{\mathcal{O}}}}\Phi(p,w)\ket{0}\,,
\end{align}
with $\Delta_{\mathcal{O}} =2+\sqrt{4+M_{\Phi}^2}$, we recast the usual result by \cite{Hofman:2008ar}:
\begin{align}
    \braket{\En{\n_1}\ldots \En{n_N}}_{\mathcal{O}(p)} = \frac{1}{(4\pi)^N} \frac{(p^2)^2}{(n_1\cdot p)^3}\ldots \frac{(p^2)^2}{(n_N\cdot p)^3}\,.
\end{align}

\section{$1/N$ corrections and the angular OPE}\label{Sec:Inter}

Energy correlators discussed so far, at leading order in $1/N$ (leading order in AdS interactions) and for states with finite $p>0$, completely factorize into the measurement of each detector separately. Indeed, in this limit the action of the energy detectors, as defined in \eqref{Eq:AdSEnOp}, simply amounts to ``counting'' AdS particle weighted by the conformally determined function of $p$ and $n$, insensitive to the detailed microscopic structure of the dynamics, such as the scaling dimensions. 

More interesting correlations arise once non-trivial $1/N$ corrections, corresponding to bulk interactions beyond just those to detector-gravitons, are taken into account. In this section we focus on the simplest class of such effects, namely non-factorized correlations at small angles between a pair of detectors (function of ``$n_1\cdot n_2$''), which are controlled by the light-ray Operator Product Expansion (OPE). 

\textit{\textbf{Small angles and light-ray OPE}}

The small-angle limit of two detectors (see, e.g., \cite{Kravchuk:2018htv, Kologlu:2019mfz}) admits a ``detector'' operator product expansion (OPE),
controlled by conformal-invariant selection rules of the form
\begin{align}\label{Eq:OPE}
    \En{\n_1}\,\En{\n_2}
    \;\sim\;
    \sum_{\tau}
    \theta^{\tau - 4}\,
    c_{\tau}^{(3)}\,
    \LO_{\tau}^{J=3}(\n)\,,
\end{align}
where $\n_1 \cdot \n_2 = \cos\theta$. The operators $\LO_{\tau}$ appearing on the
right-hand side are referred to as light-ray operators and define new ``effective detectors'' that are local in angle.\footnote{More precisely, detectors are light-ray operators inserted at spatial infinity. Furthermore, generalizations of
\eqref{Eq:OPE} include operators with transverse spin (see, e.g., \cite{Chang:2020qpj}, which do not qualitatively affect
the discussion that follows.} Such operators can be defined either as integrals of local operators along a light-ray,
as in \eqref{Eq:EnDetectorLC}, or as the outcome of an operator product expansion, as in
\eqref{Eq:OPE}. All this is true for a general CFT. Moreover when a large-$N$ expansion (its AdS dual) is available, the operators
$\LO_{\tau}$ in \eqref{Eq:OPE} can be expressed in terms of local and/or bilocal
multi-trace operators \cite{Hofman:2008ar}, as we discuss in more detail below. 

The detector $\En{\n}$ and, more generally, light-ray operators
have particularly simple transformation properties under dilatations and boosts along
the $\n$ direction, the latter usually called ``collinear spin''. Let us consider $\En{\n}$ as defined in \eqref{Eq:EnDetectorLC}. It is straightforward to
check that under dilatations
\begin{align}
    x^{\mu} \rightarrow \lambda^{-1} x^{\mu}
    \qquad\Longrightarrow\qquad
    \En{\n} \rightarrow \lambda\, \En{\n}\,.
\end{align}
Similarly, under collinear spin along the $\n$ direction,
\begin{align}\label{Eq:CollSpin}
    (\xnp , \xnm, \xnT) \rightarrow (\lambda \xnp ,\lambda^{-1} \xnm, \xnT)
    &&\Longrightarrow&&
    \En{\n} \rightarrow \lambda^{-3} \En{\n}\,.
\end{align}
Moreover, in the small-angle limit the boost acts as $\theta \rightarrow \lambda\,\theta$.\footnote{
Under boosts one may assign weights
$n^{\mu}\rightarrow \lambda n^{\mu}$ and $\bar n^{\mu}\rightarrow \lambda^{-1}\bar n^{\mu}$,
so that $\xnp = x\!\cdot\!\bar n$ and $\xnm = x\!\cdot\! n$ transform as in
\eqref{Eq:CollSpin}. Since $\theta \sim \sqrt{1-n_1\!\cdot\! n_2}$, in the collinear limit
this implies $\theta\rightarrow \lambda\,\theta$.
} As a consequence, the OPE \eqref{Eq:OPE} fixes the scaling dimension and collinear spin of the
light-ray operators $\LO_{\tau}^{J}$ to be $(\Delta_{\LO_{\tau}^J}, J_{\LO_{\tau}^J}) = (J-1,\,1-\tau-J)$.
 These transformation properties are summarized in Tab.~\ref{Tab:CFTQM}.

\begin{table}[t]
\centering
\begin{tabular}{c|c|c|c|c}
 & $\mathcal{E}$ & $\LO_{\tau}^J$ & $\theta$ & $\Lambda_{\text{IR}}$ \\
\hline
$J$ & $-3$ & $1-\tau-J$ & $1$ & $0$ \\
$\Delta$ & $1$ & $J-1$ & $0$ & $1$ \\
\end{tabular}
\caption{The collinear spin and dimension quantum numbers of operators and quantities appearing in the angular OPE \eqref{Eq:OPE}. $\LIR$ is the gap relevant for Sec.~\ref{Sec:RS}.}
\label{Tab:CFTQM}
\end{table}

In the AdS dual, the large-$N$ expansion makes it possible to identify the light-ray operator $\LO_{\tau}^{J}$ associated with the scalar field $\Phi$ introduced in \eqref{Eq:AdsScalar} with the following double-trace expression,
\begin{align}\label{Eq:DTOp}
    &\LO_{2\Delta_{\mathcal{O}}}^{J}(\n) \,=\, \lim_{\xnpd \rightarrow\infty} (\xnpd)^{2 \Delta_{\mathcal{O}}}
    \int d \xnmd \, dt \\
    &\Bigl[ (t+ i \epsilon)^{-1-J}+(-1)^J(-t+ i \epsilon)^{-1-J} \Bigr]  
     \mathcal{O}(\xnpd,\xnmd-t)\,
    \mathcal{O}(\xnpd,\xnmd+t)\,,\notag
\end{align}
where $\mathcal{O}$ is the CFT operator dual to $\Phi$ and we follow the conventions of \cite{Kravchuk:2018htv}. The $i\epsilon$ prescription in the kernel projects the bilocal operator onto a definite collinear spin $J$ and enforces positive-energy support \cite{Kravchuk:2018htv}. More generally, similarly to \eqref{Eq:DTOp} one must also consider additional double-trace light-ray operators such as those built out of the CFT stress tensor $T^{\mu \nu}$.

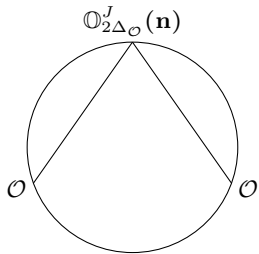
\begin{figure}
    \centering
    \begin{tikzpicture}[scale=0.7]
        \draw (0, 0) circle (2cm);

        \node at (0,2.4) {$\LO_{2 \Delta_{\mathcal{O}}}^{J}(\n)$};
        \node at (-2.2,-0.78) {$\mathcal{O}$};
        \node at (2.2,-0.78) {$\mathcal{O}$};

        \begin{feynman}
            \vertex at (-1.88,-0.68) (a1);
            \vertex at (1.88,-0.68) (a2);
            \vertex at (0,2) (c);

            \diagram*{
                (a1) -- (c),
                (a2) -- (c),
            };
        \end{feynman}
    \end{tikzpicture}
    \caption{Diagram corresponding to the matrix element in AdS one-particle state of the double-trace operator in the two-detector OPE of \eqref{Eq:OPE}. }
    \label{fig:LODoubleTrace}
\end{figure}

It is instructive to first study the action of the operator in \eqref{Eq:DTOp} at leading order in the state created by $\mathcal{O}$ with finite $p>0$. This can be computed starting directly from the ($N=\infty$) factorized form of the Witten diagram depicted in Fig.~\ref{fig:LODoubleTrace}, involving the double trace operator $\LO_{\tau}^J$. Starting from position space we get 
\begin{align}\label{Eq:DTLeadingN}
    \braket{\mathcal{O}(x) \LO_{2\Delta}^J (\n) \mathcal{O}(0)} \simeq \frac{1}{(x\cdot n)^{2\Delta+J-1}} \,,
\end{align}
where we should emphasize that it crucial to maintain the proper ordering of the operators and we are dropping overall factors. It is easy to check that the Fourier transform of \eqref{Eq:DTLeadingN} vanishes for any state with given finite $p^{\mu}=(p^0,\mathbf{0})$. Indeed we can just consider \eqref{Eq:DTLeadingN} in momentum space for a wavepacket with some smearing $\sigma$ to see that the result is exponentially suppressed,
\begin{align}
    \int d^4x e^{i p^0x^0 }  e^{-x^2 /2\sigma^2}\braket{\mathcal{O}(x) \LO_{2\Delta}^J (\n) \mathcal{O}(0)}\sim e^{-(p^0)^2\sigma^2/2}\,,
\end{align}
and vanishes when we send $\sigma \rightarrow \infty$, trying to localize the wave packet in momentum. It is easy to understand this just inspecting the associated factorized diagram in Fig.~\ref{fig:LODoubleTrace}. The only way for our particle to ``hit'' the detector is to travel directly in its direction (at the speed of light) since no interaction happens in the bulk at this order. A similar argument was used in App.~B of \cite{Hofman:2008ar} to show the $\theta$ dependence in \eqref{Eq:OPE}.

We emphasize that this is consistent with the absence of any $\theta$ dependence that we found before for states with definite $p^{\mu}$ with $p>0$ (implicitly $\sigma \rightarrow \infty$). Non-trivial angular dependence on $\n_1\cdot \n_2$, of the type seen in the OPE \eqref{Eq:OPE} arises, in states with definite $p>0$, once we include additional bulk interactions.

\textit{\textbf{One-loop correction in AdS}}

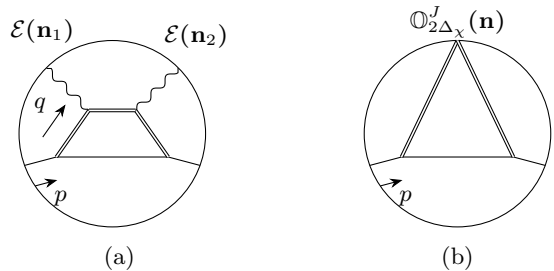
\begin{figure}[t]
\centering

\begin{minipage}[t]{0.48\columnwidth}
\centering
\begin{tikzpicture}[scale=0.62]
  \draw (0, 0) circle (2cm);

  \node at (-1.5,2.2) {$\En{\n_1}$};
  \node at ( 1.8,2.1) {$\En{\n_2}$};

  \begin{feynman}
    \vertex at (-0.5,0.5) (b1);
    \vertex at ( 0.5,0.5) (b2);
    \vertex at (-1.88,-0.68) (a1);
    \vertex at ( 1.88,-0.68) (a2);

    \vertex at (-1.41,1.4)  (c1);
    \vertex at ( 1.41,1.41) (c2);
    \vertex at (-1.2,-0.5)  (d1);
    \vertex at ( 1.2,-0.5)  (d2);

    \diagram*{
      (a1) -- [momentum'=$p$] (d1),
      (d2) -- (a2),
      (d1) -- (d2),
      (d1) -- [double, momentum=$q$] (b1),
      (d2) -- [double] (b2),
      (b1) -- [double] (b2),
      (b1) -- [photon] (c1),
      (b2) -- [photon] (c2),
    };
  \end{feynman}
\end{tikzpicture}

\vspace{0.25em}
{\small (a)}
\end{minipage}
\hfill
\begin{minipage}[t]{0.48\columnwidth}
\centering
\begin{tikzpicture}[scale=0.62]
  \draw (0, 0) circle (2cm);
  \node at (0,2.40) {$\LO_{2 \Delta_{\chi}}^{J}(\n)$};

  \begin{feynman}
    \vertex at (-1.88,-0.68) (a1);
    \vertex at ( 1.88,-0.68) (a2);

    \vertex at (-1.2,-0.5) (d1);
    \vertex at ( 1.2,-0.5) (d2);
    \vertex at (0,2) (c);

    \diagram*{
      (a1) -- [momentum'=$p$] (d1),
      (d2) -- (a2),
      (d1) -- (d2),
      (c) -- [double] (d1),
      (d2) -- [double] (c),
    };
  \end{feynman}
\end{tikzpicture}

\vspace{0.25em}
{\small (b)}
\end{minipage}

\caption{(a) A $1/N$ correction to the two-point EC where non-trivial angular dependence first appear for our state. (b) The leading OPE contribution to (a) as in \eqref{Eq:OPE}.}
\label{fig:Loop}
\end{figure}

For our AdS single-particle state these bulk interactions appear at one loop order. To simplify the discussion, it is easier to have two separate species of AdS particles by introducing another scalar $X$ in \eqref{Eq:AdsScalar}, with action
\begin{align}
    S_{X}=\int d^4xdw\sqrt{G} \left(\frac{1}{2} \partial_MX\partial^{M}X -\frac{1}{2}M_X^2 X^2 -\frac{\lambda}{3!} X \Phi^2 \right)\,.
\end{align}
We consider the loop diagram shown in Fig.~\ref{fig:Loop}~(a), where double lines
denote the field $X$ and single lines denote $\Phi$. While the full expression for this diagram
is rather involved, to highlight the physics of interest it is useful to focus
on the integration region where $q\sim 1/r$ (with $r$ the detector size).

Indeed, in this regime the loop integral can retain a non-analytic
angular dependence. We can extract the small-angle behavior following our
previous considerations. In the small-angle limit, the OPE in \eqref{Eq:OPE} applies, and the diagram in
Fig.~\ref{fig:Loop}~(a) effectively collapses to that in
Fig.~\ref{fig:Loop}~(b), i.e.
\begin{align}\label{Eq:OPEFull}
    \braket{\En{\n_1} \En{\n_2}}_{\mathcal{O}(p)} \underset{\theta \ll 1}{\longrightarrow} &\lambda^2\theta^{2\Delta_X - 4}\braket{\LO_{2\Delta_X}^{J=3}}_{\mathcal{O}(p)}\\=&\lambda^2\frac{(p^2)^4}{(n_1\cdot p)^6}
 \theta^{2\Delta_X - 4} \frac{(p^2)^{\Delta_X-2}}{(p \cdot n_1)^{2\Delta_X-4}}\,, \notag
\end{align}
where we are dropping again $O(1)$ factors. In the second line, we computed the expectation value $\braket{\LO_{2\Delta_{X}}^{J=3}}_{\mathcal{O}(p)}$, depicted in Fig.~\ref{fig:Loop}, by a direct calculation. Furthermore it can be checked that for even collinear spin $J$, \eqref{Eq:DTOp} collapses to a light-ray integrated local operator whose expectation value $\braket{\LO^{J}_{2\Delta_{\chi}}}_{\mathcal{O}(p)}$ is now just determined by conformal invariance. The analytic continuation of this to $J=3$ is consistent with \eqref{Eq:OPEFull}.
 
The result in \eqref{Eq:OPEFull} illustrates how measurement of the small angle $\theta$ dependence of the energy correlators, allows us to extract CFT scaling (dual to AdS) masses.\footnote{More precisely the OPE of \eqref{Eq:OPE} incudes all light-ray operators with spin $J=3$, including double trace operators built out of two stress-tensors, with integer twist leading to analytic $\theta^2$ dependence at leading order in $1/N$. Even if this power of $\theta$ dominates that in \eqref{Eq:OPEFull}, in principle the non-analytic $\theta$ dependence in \eqref{Eq:OPEFull} can be used to extract $\Delta_{\chi}$.} 

\section{Flowing towards the CFT}\label{Sec:IrrDef}
We now consider a CFT deformed by an irrelevant interaction
\begin{align}\label{Eq:CFTBreak}
        \mathcal{L}(\Lambda_{\text{UV}}) = \mathcal{L}_{\text{CFT}} + g_0  \mathcal{O}_{\Psi}\,,
\end{align}
where $\mathcal{O}_{\Psi}$ is an operator of scaling dimension $\Delta_{\Psi}>4$. The effect of CFT breaking can be modeled via AdS/CFT by considering a bulk-scalar $\Psi$ with mass $M_{\Psi}^2 =\Delta_{\Psi} (\Delta_{\Psi}-4)$. The EFT cut-off is modeled by a UV boundary at a location $w_0 = 1/\Lambda_{\text{UV}}$ and $g_0$ in \eqref{Eq:CFTBreak} is imposed as a boundary condition:
\begin{align}\label{Eq:PsiSol}
    \Psi(w)
    =
    g_0 \left( \frac{w_0}{w} \right)^{\Delta_{\Psi}-4}\,.
\end{align}
This profile back-reacts on the geometry, which can be parametrized as
\begin{align}\label{Eq:GenMetric}
    ds^2
    =\bar{G}_{MN}dX^M dX^N=
    \frac{\rho(w)\, dx^2 - dw^2}{w^2}\,,
\end{align}
where $\rho(w)$ is determined by solving the Einstein equations
$\mathcal{G}^{MN}=\mathcal{T}^{MN}_{\Psi}$, with $\mathcal{G}^{MN}$ the Einstein tensor
and $\mathcal{T}^{MN}_{\Psi}$ the stress tensor sourced by the $\Psi$ profile
\eqref{Eq:PsiSol}.

Expanding to first non-trivial order in $g_0$, the independent component of the
Einstein equations reduces to
\begin{align}\label{Eq:SolBack}
        3\frac{\rho'}{w}
        =
        -\frac{g_0^2}{w_0^2}
        \left(\frac{w_0}{w}\right)^{2(\Delta_{\Psi}-4)}
        (\Delta_{\Psi}-4)\,,
\end{align}
where derivatives are with respect to $w$. Solving this equation yields the approximate metric factor
\begin{align}\label{Eq:ApproxSolutionMetric}
    \rho(w)
    =
    1
    - \frac{g_0^2}{6}
    \left(\frac{w_0}{w}\right)^{2(\Delta_{\Psi}-4)}
    + \mathcal{O}(g_0^4)\,.
\end{align}
While this is an approximate solution, any Lorentz-invariant irrelevant deformation generates a metric of the form of \eqref{Eq:GenMetric}, where $\rho(w) \xrightarrow[w\to\infty]{} 1$, i.e. asymptotically AdS. 

We now consider our scalar of \eqref{Eq:AdsScalar} in this more general metric. We start with one detector, closely following the discussion of the diagram in \eqref{Eq:OnePointDiagram}. 

It is convenient to decompose the operator corresponding to the external states as
\begin{align}
    \Phi(x,w) = \!\! \int_{0}^{\infty}\!\!\!\! dm \,\mathcal{N}_{m}\!\!\!\int \!\!d\Omega_{\mathbf{p}} f_m(w)\left( a_{m}^{\dagger}(\mathbf{p}) e^{i px}+h.c.\!\right)\,,
\end{align}
with $f_m(w)$ the mode function:
\begin{align}
    -\frac{w^5}{\rho(w)^2} \partial_w \!\!\left(\!\! \frac{\rho(w)^2}{w^3}\partial_w f_{m}(w)\!\! \right)\!\!+\!M_{\Phi}^2 f_{m}(w) \!=\! \frac{w^2}{\rho(w)} m^2 f_{m}(w)\,,
\end{align}
and $\mathcal{N}_m$ a normalization factor, together with the simple choice of the boundary condition $f_{m}(w_0) = 0$. From the previous equation we can extract the following asymptotic solution for the mode function
\begin{align}
    f_{m}(w) \xrightarrow[w\to\infty]{} w^2\sqrt{\frac{2}{\pi}}\frac{\cos(mw+\phi)}{\sqrt{mw}}\left(1+ O(1-\rho(w)) \right)\,.
\end{align}
The leading behavior corresponds to the plane wave we found in \eqref{Eq:AdsPlaneWave}.

Next, we need to consider the bulk-to-boundary propagator of the graviton $K_{\mu \nu, \alpha \beta}(q,w)$ to plug in \eqref{Eq:1PointFull}. To explain the relevant physics it is convenient, momentarily, to pretend the graviton is actually itself a scalar and consider $K(q,w)$, its bulk-boundary propagator. Later on we will be more precise. We are interested again in the limit $q^{\mu}\sim 1/w \sim 1/\xnp$ where $\xnp$ goes to infinity. $K(q,w)$ can be found by solving the EoM that takes the form 
\begin{align}\label{Eq:ScalarPMetric}
    \left(-\frac{w^5}{\rho(w)^2} \partial_w  \frac{\rho(w)^2}{w^3}\partial_w  +M^2- \frac{w^2}{\rho(w)} q^2 \right)K(q,w)=0\,.
\end{align}

We can notice that in the limit we are interested in, $K(q,w)$ still asymptotes to the pure AdS solution,\footnote{Strictly speaking there are two AdS solutions, however when $K(q,w)$ is evaluated for spacelike momenta, see \eqref{Eq:LCIntgral}, only one is acceptable and doesn't explode for large $w$.} i.e. 
\begin{align}\label{Eq:AsPropagatorAdS}
    K(q,w) \sim c K_{\text{AdS}} (q,w) \left( 1+ \mathcal{O}(\rho(w)-1)\right)\,,
\end{align}
possibly up to a proportionality constant $c$.
Essentially the same logic applies to the bulk-to-boundary propagator of the graviton which asymptotes to its AdS form up to a normalization factor $c$. The rest of the calculation parallels the AdS case and we get
\begin{align}\label{Eq:OnePointAdsDef}
    \frac{\braket{\Phi(p,w_f)|\En{\n}|\Phi(p,w_i)}}{\braket{\Phi(p,w_f)|\Phi(p,w_i)}} = c\frac{1}{4 \pi} \frac{(p^2)^2}{(n \cdot p)^3}\,.
\end{align}
The normalization factor $c$ can be fixed by Energy conservation which requires $c=1$.

We emphasize that the same result can be analogously obtained by treating the state in the Point-Particle approximation, in the same fashion as Sec.~\ref{Sec:AdSPP}. Indeed, also in the metrics \eqref{Eq:PsiSol} and \eqref{Eq:ApproxSolutionMetric}, the particle continues to propagate to large $w$ and eventually asymptotes to the speed of light, approaching the classical AdS trajectory
$x^{\mu}_{\text{PP}}(w) \sim p^{\mu}\, w/p $. From the discussion above, in particular around \eqref{Eq:AsPropagatorAdS}, it then follows directly that one recovers the result in \eqref{Eq:EPPresult}, which is completely analogous to \eqref{Eq:OnePointAdsDef} upon setting $c=1$. 

Again, we can consider multiple detectors and find the same results holds as in exact AdS
\begin{align}\label{Eq:NDetectorAsAdS}
    \frac{\braket{\Phi(p,w_f)|\En{\n_1}\ldots\En{\n_N}| \Phi(p,w_i)}}{\braket{\Phi(p,w_f)|\Phi(p,w_i)}}
    = \prod_i\frac{1}{(4 \pi)} \frac{(p^2)^2}{(n_i \cdot p)^3}\,.
\end{align}
In other words, the action of the Energy detector is given by \eqref{Eq:AdSEnOp}. Again this depends on the vanishing of diagrams analogous to Fig.~\ref{fig:PP_Two} (b) and (c), which we turn to demonstrate now.

\textbf{\textit{Factorization in the (deformed) CFT}}

Let us start with a more detailed derivation of \eqref{Eq:OnePointAdsDef}. The insertion of the detector $\En{\n}$ on the boundary corresponds to a perturbation of the metric,
\begin{align}\label{Eq:MetricPerturbed}
    ds^2 \rightarrow ds^2 + (\xnpd)^2 \delta(\xnp - \xnpd)\, h(\xnT,w)\, d\xnp d\xnp\,,
\end{align}
where $\xnpd$ is eventually sent to infinity. This form of  perturbation is a consequence of the source being integrated along a light-ray. This follows directly from the discussion around \eqref{Eq:LCIntgral}, which does not rely on the AdS geometry but only on the Lorentz structure of the bulk-to-boundary propagator $K_{\mu \nu,\alpha \beta}$. The precise form of $h(x^{T}_{\n},w)$ can be found by solving the Einstein equations $\mathcal{G}^{MN} = \mathcal{T}_{\Psi}^{MN}$ for the metric in \eqref{Eq:MetricPerturbed}. The $M=N=5$ component coincides with \eqref{Eq:SolBack},  while using 
\begin{align}
    \mathcal{T}_{\Psi}^{\mu  \nu} = -\frac{1}{2}G^{\mu \nu}\left(\bar{G}^{55}(\partial_{w}\Psi(w))^2 -  M_{\Psi}^2 \Psi(w)^2\right)\,,
\end{align}
with $\Psi(w)$ the classical profile in \eqref{Eq:PsiSol} and $G_{\mu \nu}$ the metric in \eqref{Eq:MetricPerturbed}. We get
\begin{align}\label{Eq:EoMGraviton}
   \left(
   \frac{w^3}{\rho(w)}\,\partial_w \frac{\rho(w)^2}{w^3}\,\partial_w
   + (\nabla_{\n}^T)^2
   \right)
   \frac{w^2}{\rho(w)}\, h(x^{T}_{\n},w)
   = 0 \, .
\end{align}
In the limit $w \rightarrow \infty$, \eqref{Eq:EoMGraviton} asymptotes to the AdS equation of motion, insensitive to the deformation:
\begin{align}
    h (x_{\n}^T,w)\xrightarrow[w\to\infty]{}\int\!\! \frac{d^2 q_{\n}^T}{(2 \pi)^2}\,
    e^{i q_{\n}^T\cdot x_{\n}^T}
    \frac{(q_{\n}^T)^2}{2}\BK{2}(w\sqrt{(q_{\n}^T)^2})\,.
\end{align}
as in \eqref{Eq:LCIntgral}.
This clarifies the derivation of \eqref{Eq:OnePointAdsDef} (with $c=1$ by energy conservation). 

Before turning to multiple detectors it is worth noticing that \eqref{Eq:EoMGraviton} coincides with the equation of motion of a massless ``scalar field''
 $\equiv\frac{w^2}{\rho(w)}\, h(x^{T}_{\n},w)$,
in accordance with \eqref{Eq:ScalarPMetric}, upon setting $q^{+}_{\n}=0$ as enforced by the light-ray integration $\int d \xnmd$. Furthermore, as an aside, it is remarkable that \eqref{Eq:ScalarPMetric} is exactly linear in $h$. 
In other words, even in this more general background, the detector insertion is dual to a geometry 
that solves the Einstein equations exactly at linear order, just as in the case of AdS shockwave 
geometries corresponding to CFT detector operators.

Let us now turn to multiple detector insertions. It is convenient to describe the state as a PP moving as
$x^{\mu}_{\text{PP}}(w) \sim p^{\mu} w / p$ for large $w$. 

We consider, for simplicity, a two-point energy correlator. One can write
\begin{align}
    \braket{\En{\n_1}\En{\n_2}}_{\text{PP}}
    \!\!=\!\!
    \braket{\En{\n_1}}_{\text{PP}}\!\braket{\En{\n_2}}_{\text{PP}}
   \!\! +\!
    \braket{\En{\n_1}\En{\n_2}}_{\text{PP}}^{\text{n.f.}}  ,
\end{align}
where the first term corresponds to diagrams with the topology of Fig.~\ref{fig:PP_Two}(a) and factorizes in complete analogy with the pure AdS case. The second term (non factorized) receives potential contributions from diagrams analogous to Fig.~\ref{fig:PP_Two}~(b) and (c). We now show that these two contributions vanish.\footnote{
In pure AdS, as argued in \cite{Hofman:2008ar}, it is possible to perform a global change of
coordinates, schematically ``$x \rightarrow y$'', such that all detectors lie on a single
null surface (for a short review of this point, see, e.g., \cite{Belin:2020lsr}).
In these coordinates the metric takes the form
\begin{align}
    ds^2_{\text{AdS}}
    \;\rightarrow\;
    ds^2_{\text{AdS}}
    + \delta(y^+)\left(h_1(y_T,w)+\ldots+h_N(y_T,w)\right)\, dy^+ dy^+\,,
\end{align}
with $h_1,\ldots,h_N$ related to the individual detector insertions. The equation of motion
for the full perturbation $h_1+\ldots+h_N$ in these coordinates is of the form
\eqref{Eq:EoMGraviton} and is exactly linear in $h_1+\ldots+h_N$, so that no nonlinearities
emerge.

By contrast, before performing the change of coordinates, while
\eqref{Eq:EoMGraviton} is exactly linear for a single detector insertion, it is not linear
for a superposition of insertions. The resulting nonlinearities are localized at the
coincident support of the detectors and are proportional to
$\delta(\hat{x}_{\n_1}^+ - x_{\n_1}^+)\,\delta(\hat{x}_{\n_2}^+ - x_{\n_2}^+)$.
}

Let us begin with Fig.~\ref{fig:PP_Two}(c), where the point-particle interacts directly with both detectors. The two light-ray integrals enforce the conditions
\[
x_{\text{PP},\n_1}^+ = \hat{x}_{\n_1}^+ \,, \qquad
x_{\text{PP},\n_2}^+ = \hat{x}_{\n_2}^+ \,,
\]
which imply
\begin{align}
    w \;=\; p\,\frac{\hat{x}_{\n_1}^+}{p_{\n_1}^+}
    \qquad\text{and}\qquad
    w \;=\; p\,\frac{\hat{x}_{\n_2}^+}{p_{\n_2}^+}\, .
\end{align}
It is clear that, when the limits are taken sequentially, these two conditions cannot be satisfied simultaneously. After taking the limit $\hat{x}_{\n_1}^+ \to \infty$, the first constraint forces $w \to \infty$ for any $p>0$, while the second constraint keeps $w$ finite as long as $\hat{x}_{\n_2}^+$ is held fixed. Consequently, the two constraints are incompatible for any $p>0$, and the contribution vanishes.

We now turn to Fig.~\ref{fig:PP_Two}(b). The two detectors interact at the same bulk point
$Y^M=(y^{\mu},w)$, such that
\begin{align}
    y_{\n_1}^+ = \hat{x}_{\n_1}^+\,, 
    \qquad
    y_{\n_2}^+ = \hat{x}_{\n_2}^+ \,,
\end{align}
as enforced by the two $\delta$ functions arising from the corresponding light-ray
integrals as in \eqref{Eq:MetricPerturbed}. We consider the internal graviton propagating between $(y,w)$ and a point
$(x_{\text{PP}}^{\mu}(w'),w')$ on the classical trajectory of the point-particle,
\begin{align}\label{Eq:Bulk-to-Bulk-graviton}
    G_{\alpha\beta, \gamma \delta}(y-x_{\text{PP}}(w'),w,w')\,,
\end{align}
where we keep only four-dimensional indices and work in axial gauge. Focusing on the
three-graviton vertex of the schematic form
$h_{\alpha}^{\beta} h_{\beta}^{\gamma} h_{\gamma}^{\alpha}$
(the discussion for other Lorentz structures is analogous), we obtain
\begin{align}\label{Eq:TwoPoint2b}
    \braket{\En{\n_1}\En{\n_2}}^{\text{2b}}\!\!\!
    \sim\!
    (\hat{x}_{\n_1}^+\hat{x}_{\n_2}^+)^2\!\!\!
    \int \frac{d^2 y^T\, dw}{\sqrt{\bar{G}}}\,
    h(y_{\n_1}^T,w)\, h(y_{\n_2}^T,w)
     \notag\\
    \int\!\! \frac{dw'}{(w')^5}\,
    \bar{n}_1^{\alpha}\bar{n}_2^{\beta}\,
    G_{\alpha \beta, \gamma \delta}(y-x_{\text{PP}}(w'),w,w')\,
    \mathcal{T}_{\text{PP}}^{\gamma \delta}(x_{\text{PP}}(w'),w'),
\end{align}
where the $d^2y^T$ integration is over the plane orthogonal to both $n_1$ and $n_2$, $\mathcal{T}_{\text{PP}}^{\gamma \delta}$ is the point-particle stress tensor, with
indices contracted using the bulk metric in \eqref{Eq:GenMetric}, and where the limit
$\hat{x}_{\n_1}^+,\hat{x}_{\n_2}^+\rightarrow\infty$ is understood. We now define $r \;\sim\; \hat{x}_{\n_1}^+ \;\sim\; \hat{x}_{\n_2}^+$ as our large expansion parameter, which is eventually sent to infinity, and study the
scaling of the integral in the various regions obtained by rescaling the integration
variables with $r$.

First, when $w$ and $w'$ are kept fixed and the separation
$(y-x_{\text{PP}}(w'))^2 \sim r^2$, the bulk-to-bulk graviton propagator approximate a boundary-to-boundary propagator up to subleading corrections, behaving as
\[
G_{\alpha \beta, \gamma \delta} \;\sim\; \frac{1}{r^8}\left(1+\mathcal{O}(\lambda^2)\right)\,.
\]
It is easy to check that the measure in \eqref{Eq:TwoPoint2b} is not enough to overcome this suppression.

The leading contribution instead arises when all integration variables scale with $r$.
In this regime the propagator scales as
$G_{\alpha \beta, \gamma \delta} \sim 1/r^4$, while each contraction with $\bar n_i^\mu$
produces a factor of order $r^2$. Moreover, the point-particle stress tensor behaves as $\mathcal{T}_{\text{PP},\gamma \delta}(r,r) \sim r^3$ (see \eqref{Eq:StressTensorMassles}) and $h(r,r)\sim 1/r^4$. Putting all factors together, we
find
\begin{align}
    \braket{\En{\n_1}\En{\n_2}}^{\text{2b}} \;\sim\; \frac{1}{r^2}\,,
\end{align}
which vanishes in the large-$r$ limit. 

Therefore we conclude that, in $N\rightarrow\infty$, factorization holds in the CFT (agreeing with \cite{Hofman:2008ar}) as well as in its irrelevant deformation.

\section{Introducing a gap with an IR boundary (RS1)}\label{Sec:RS}
So far, we have considered gapless theories dual to asymptotically $\text{AdS}_5$. As already emphasized, measurements of energy flow are made in the far IR, at arbitrarily large distances from the interaction point. In the AdS description,
this point was already discussed around \eqref{Eq:LCIntgral}, where we argued
that the detector effectively measures the point-particle at positions
$w \sim r$, with $r$ the detector size. This is the key to why, in the absence of bulk interactions, ECs in the geometry of \eqref{Eq:GenMetric} are the same as in pure AdS.

However, any geometry that departs significantly from AdS in the deep IR,
i.e.\ for $w \sim r$, will generically lead to results that differ from those
discussed so far. For instance, a relevant deformation, such as the one
considered in Sec.~\ref{Sec:IrrDef} but when $\Delta_{\Psi}<4$, poses a potential problem: at large $w$ the
metric backreaction in \eqref{Eq:ApproxSolutionMetric} grows and perturbative
control of the geometry is eventually lost, precisely in the large $w$ region relevant for the energy measurements.

The simplest geometry that departs from AdS in the infrared is the truncation of AdS by a ``IR boundary''. This is the scenario familiar from RS1
 \cite{Randall:1999ee}, described by the metric
\begin{align}\label{Eq:RSMetric}
   ds^2
   =
   \frac{dx^2 - dw^2}{w^2}\,,
   \qquad
   0 < w < w_{\text{IR}} \equiv \frac{1}{\LIR}\,.
\end{align}
See \cite{Randall:1999ee} for the proper treatment of the IR boundary condition for the gravitational field. This gap represents a confinement scale $\LIR$ from the dual 4D viewpoint (see, e.g., \cite{Maldacena:2003nj}). 

We now reconsider the scalar field $\Phi$ introduced in
\eqref{Eq:AdsScalar}, but propagating in the background
\eqref{Eq:RSMetric}. As well-known, the IR boundary results in a discrete Kaluza-Klein spectrum, with 4D mass gap $\LIR \sim 1/w_{\text{IR}}$. If this gap is much smaller than collider energies, $\LIR \ll p$, one expects the gap to be negligible. However, this appears to conflict with our earlier analysis in AdS, where we saw that the interaction of the PP with the detector (graviton) takes place in the asymptotic IR, $w \sim r \rightarrow \infty$. But if the confinement length scale is smaller than the detector radius, $w_{\text{IR}} < r$, this interaction point is clearly absent in the truncated bulk. Therefore it would appear that the gap/confinement has a major impact on the energy correlators, in contradiction to our expectation.

The situation is intuitively clear from the dual 4D viewpoint. In the CFT limit (no gap), the CFT state produced at the collision point can grow to a size of the detector radius $r$, and therefore give a non-trivial energy measurement even if its center-of-mass is not headed in the direction of that detector. But with finite confinement length, the state produced at the collision is a single stable confined hadron (Kaluza-Klein resonance) in the large-$N$ limit and will not be detected unless its center-of-mass is exactly in the direction of the detector. However, we will argue that $1/N$ (bulk interaction) corrections and the accompanying hadron decays can dramatically change this conclusion, with the ECs being well-approximated by the gapless limit as first intuited. 

\emph{\textbf{Free stable KK-excitations/hadrons}}

We first study energy correlators in the gapped theory in the absence of bulk interactions ($N = \infty$), showing the sharp departure from AdS. 

The continuum decomposition
\eqref{Eq:ContKKDecomposition} is replaced by a discrete mode expansion,
\begin{align}\label{Eq:RSKKDecomp}
    \Phi(x,w)
    =
    \sum_a \mathcal{N}_a
    \int d \Omega_{\mathbf{p}}^{m_a}\,
    f_{a}(w)
    \left(
        a^{\dagger}_{m_a}(\mathbf{p})\, e^{i p\cdot x}
        + \text{h.c.}
    \right)\,,
\end{align}
where $m_a$ denotes the discrete mass spectrum,
$m_a \sim a \pi \, \Lambda_{\text{IR}}$ and for simplicity we impose  Dirichlet boundary conditions on $\Phi$. 
The corresponding mode functions are
\begin{align}
    f_a(w) = w^2 \, J_{\nu}(m_a w)\,,&&f_{k}(w_{\text{IR}})=0\,,
\end{align}
with normalization coefficients $\mathcal{N}_a$.\footnote{
In the limit $\Lambda_{\text{IR}} \to 0$, the discrete sum reproduces the
continuum decomposition of \eqref{Eq:ContKKDecomposition}: $\sum_a \mathcal{N}_a
\;\underset{\Lambda_{\text{IR}}\rightarrow 0}{\longrightarrow}\;
\int dm \,\sqrt{m}$.
}
The creation and annihilation operators satisfy the standard 4D canonical
commutation relations.
As a result, the Feynman bulk-to-bulk propagator in
 \eqref{Eq:Bulktobulkpropagator} reduces to
\begin{align}
    G_F(p,w_f,w_i)
    =
    i \sum_a \left( \mathcal{N}_a\right)^2
    \frac{
        f_a(w_f)\,
        f_a(w_i)
    }{
        p^2 - m_a^2 + i \epsilon
    }\,.
\end{align}

The calculation of ECs closely parallels the case of a four-dimensional free particle (see, e.g., Ref.~\cite{Bauer:2008dt}) as we now outline. It is convenient to treat $\En{\n}$ directly as an operator in the
second-quantized theory, as in \eqref{Eq:LocalStressTensor},
\begin{align}\label{Eq:DetRS}
    \mathcal{E}(\n)
    &=
    \lim_{r\rightarrow\infty} r^2
    \int_0^{+\infty} dt\\
    &\int \frac{dwd^4 y}{w^5}\,  \,
    \n_i\,
    K_{0i,\alpha\beta}\bigl((t,r\n)-y,w\bigr)\,
    \mathcal{T}_{\Phi}^{\alpha \beta}(y,w)\,. \notag
\end{align}
Notice that, in the expression above, we
have used the original definition of the energy detector
\eqref{Eq:EnDetector}, since in the present gapped setup, with a finite maximal boost, it is unclear whether one can push the detector to null infinity and employ the light-cone definition \eqref{Eq:EnDetectorLC}. 
Indeed,  Ref.~\cite{Csaki:2024joe} takes the light-cone limit of detectors even in the gapped context, which leads to qualitatively different results from ours. For instance, in the leading approximation (no bulk interactions beyond those with the detector gravitational field) we will show that ECs reduce to those of {\it free} 4D KK particles. In this we disagree with \cite{Csaki:2024joe} which claims a non-trivial angular correlation between detectors. 

The action of the detector \eqref{Eq:DetRS} can be obtained by expanding the
stress tensor $\mathcal{T}_{\Phi}^{\alpha \beta}$ in terms of creation
and annihilation operators, using the mode expansion
\eqref{Eq:RSKKDecomp}. We obtain
\begin{align}\label{Eq:RSDetSecondQ}
    &\En{\n}
    =
    \lim_{r\rightarrow\infty} r^2
    \int_0^{\infty} dt
    \sum_{a,b}
    \int d\Omega_{\mathbf{p}}^{m_a}
    d\Omega_{\mathbf{k}}^{m_b}
    \notag\\
    &
    \Bigl[
        g_{ab}(p,k)\,
        a_{m_a}(\mathbf{p})
        a_{m_{b}}^{\dagger}(\mathbf{k})
        e^{-i (t,r\n)\cdot (p-k)}
        + (k\rightarrow -k)
        + \text{h.c.}
    \Bigr]\,,
\end{align}
where $g_{ab}(p,k)$ encodes the bulk-to-boundary graviton propagator convoluted with
the mode functions $f_m(w)$.

The angular integrals can
be evaluated using a stationary-phase approximation in the large-$r$ limit, which enforces
\begin{equation}
    \mathbf{p} \parallel \mathbf{k} \parallel \n \,.
\end{equation}
As a result, we find
\begin{align}
    \En{\n}
    &=
    \frac{1}{(2\pi)^4}
    \lim_{r\rightarrow\infty}
    \int_0^{\infty} dt
    \sum_{a,b}
    \int
    \frac{d|\mathbf{p}|^2}{2\omega_{\mathbf{p}}}
    \frac{d|\mathbf{k}|^2}{2\omega_{\mathbf{k}}}
    \notag\\
    &\quad
    \Bigl[
        g_{ab}(p,k)\,
        a_{m_a}(\mathbf{p})
        a_{m_{b}}^{\dagger}(\mathbf{k})
        e^{-i t(\omega_{\mathbf{p}}-\omega_{\mathbf{k}})
          + i r(\mathbf{p}-\mathbf{k})}
        + \ldots
    \Bigr]\,.
\end{align}

We now switch to light-cone coordinates. In the large-$r$ limit, the time integral becomes
\begin{align}
    \lim_{r\rightarrow \infty}
    \int_0^{\infty} dt\,
    e^{i (\omega_{\mathbf{p}}-\omega_{\mathbf{k}})t
      - i (|\mathbf{p}|-|\mathbf{k}|)r}
   &=
    \lim_{r \rightarrow \infty}
    \int_{-r}^{\infty} d\xnm\,
    e^{i (\omega_{\mathbf{p}}-\omega_{\mathbf{k}})\xnm}
    \notag\\
    &=
    \delta(\omega_{\mathbf{p}}-\omega_{\mathbf{k}})\,
    e^{i r (p_{\n}^--{k}_{\n}^-)}\,,
\end{align}
which impose that no energy so that no further integral remains. We are left with
\begin{align}
    \En{\n}
    &=
    -\sum_{a,b}
    \lim_{r\rightarrow \infty}
    \int
    \frac{d|\mathbf{p}|\,|\mathbf{p}|}{2 \omega_{\mathbf{p}}}
    \frac{d|\mathbf{k}|\,|\mathbf{k}|}{2 \omega_{\mathbf{k}}}
    \notag\\
    &\quad\times
    \Bigl[
        g_{ab}(p,-k)\,
        a_{m_a}(\mathbf{p})
        a_{m_{b}}^{\dagger}(\mathbf{k})
        e^{-i (p_{\n}^- - {k}_{\n}^-) r}
        + \text{h.c.}
    \Bigr]\,,
\end{align}
where, again, $\omega_{\p} = \omega_{\mathbf{k}}$.

For $a\neq b$, the phase oscillates rapidly as $r\to\infty$. When convoluted
with any smooth test function, these contributions vanish, and only the
diagonal terms with $n=n'$ survive. In this case we have 
\begin{align}
     g_{aa}(p,-p) = \left(\mathcal{N}_{a}\right)^2 \omega_{\p} |\p|\int \frac{dw}{w^3} \left(f_{a}(w)\right)^2  =  \omega_{\p} |\p|\,,
\end{align}
where we used that modes are normalized and that $K_{0 i,\alpha\beta}(q,w)\xrightarrow[q\to 0]{}\frac{1}{w^2}
\eta_{0\alpha}\eta_{i\beta}
$.

Putting all together,  we finally obtain
\begin{align}
    \En{\n}
    =
    \sum_a
    \int d\Omega_{\mathbf{p}}^{m_a}
    \,
    \omega_{\mathbf{p}}\,
    a_{m_a}^{\dagger}(\mathbf{p})
    a_{m_a}(\mathbf{p})\,
    \delta^{(2)}(\n-\hat{\mathbf{p}})\,,
\end{align}
where $\delta^{(2)}(\n-\hat{\mathbf{p}})$ is the obvious $\delta$-function in angle. In other words, the energy detector simply counts the energy carried by modes
propagating in the direction $\n$. This result matches our earlier physical
interpretation: in the limit $r\rightarrow\infty$, the KK-excitations/hadrons behave as
effectively point-like, free particles.

\textit{\textbf{Beyond stable KK-excitations/hadrons}}

Once we include perturbative bulk interactions, the KK excitations will be able to decay, cascading down to the lightest KK modes (before decaying back to the standard model, which we do not model here). 
The process starts with the hard collision which creates a KK excitation with 4D mass $p$. Each decay process will typically result in KK decay-products which have smaller but comparable 4D mass to the parent KK excitation. Therefore, for $p \gg \LIR$, the entire cascade process will not be captured by any fixed order in the bulk perturbation expansion. Furthermore, we are trying to use AdS EFT above $\LIR$, so we are assuming that the dual (approximate) CFT has a large gap in the spectrum of scaling dimensions and that we have integrated out the AdS particles dual to the operators with large dimensions. However, even operators with large dimensions, ${\cal O}_{\Delta \gg 1}$, can interpolate states with 4D masses $\ll p$, and therefore arise in the cascade decays. These therefore cannot be captured by the bulk EFT. For both these reasons, it appears that even the weakly coupled RS1 effective description does not allow one to compute exclusive processes at high energies $p$. Energy correlators (at large angles) are partially inclusive enough to evade this dismal conclusion because the energy flow is determined {\it before} confinement effects take place and therefore approximate the correlators of the gapless (full AdS) theory, where fixed-order AdS perturbation theory is valid. 

An analogous situation arises in large-$N$ QCD with  a small number of massless quarks, in which the 
4D spectrum consists of 
infinitely-many stable mesons at $N= \infty$, which then 
decay when we include $1/N$ corrections, with narrow widths of order 
$\Gamma_{\text{meson}}/m_{\text{meson}} \sim 1/N$. This seems very different from the perturbative QCD picture of gapless states of quarks and gluons. And yet, for 
 sufficiently excited mesons, $m_{\text{meson}} > N \Lambda_{\text{QCD}}$, the mesons are as wide as their typical $\Lambda_{\text{QCD}}$ splittings and one thereby returns to a continuous spectrum in this regime. In this regime, 
$p > N \Lambda_{\text{QCD}}$, perturbative QCD is expected to predict energy correlators accurately, at least for $\sim O(1)$ angles. In the case of the 4D dual confining theory to the RS1 truncated bulk, the discrete resonances again effectively return to a continuum in the regime 
$m_n > N \LIR$ for the same reason. In this regime, 
$p > N \LIR$, the gapless CFT accurately predicts the energy correlators, again at least for $\sim O(1)$ angles, as we explain below. 

This also follows from a real-time argument. The collision produces a high-mass single-trace ``meson'' state which expands close to the speed of light. This process is very well approximated by the exact CFT until a time of order $1/\LIR \sim w_{IR}$ when confinement effects become important. However, if this state splits (``decays") into many single-trace states
before this confinement time, confinement will greatly affect the individual low-lying single-trace ``mesons'' but this will not greatly affect the overall energy flow, and therefore agree with CFT predictions for energy correlators. This ``decay''-time is of order $N/p$, so the CFT predictions should be valid for 
$p \gg N \LIR$. 

We now study the systematic corrections to the CFT predictions at small angles due to the gap.

\textit{\textbf{Small angle sensitivity to confinement}}

While in the $p \gg N \LIR$ regime the energy correlators are well approximated by the CFT
(full AdS) results, deviations from the CFT behavior are expected in special configurations, such as detectors placed at very small angular separation.
In this section we argue that precisely in this regime these corrections can be estimated
and the angular dependence determined as a consequence of (nonlinearly realized) conformal
invariance. 

At first sight, it may appear that since the IR boundary breaks AdS isometries, conformal-invariance control of the OPE, introduced in Sec.~\ref{Sec:Inter}, is lost. However, there is a limit of the IR boundary set-up, realized in RS1 \cite{Randall:1999ee}, in which this breaking is dual to {\it spontaneously} broken conformal invariance \cite{Rattazzi:2000hs}. The associated Nambu-Goldstone boson (the ``dilaton'') is given in the RS1 dual by the radion field, whose fluctations non-linearly realize conformal invariance. 
Therefore, in such a set-up we can continue to use conformal symmetry to control the angular OPE, where the canonical radion VEV, effectively $\LIR$, is treated as a dimension-$1$ spurion of conformal breaking, see Tab.~\ref{Tab:CFTQM}.  

Thus in \eqref{Eq:OPE}, in addition to the OPE terms insensitive to $\LIR$, there will be small corrections proprtional to powers of $\LIR$,
\begin{align}\label{Eq:OPEIR}
    \En{\n_1}\,\En{\n_2}
    \;\supset \sum_{\tau}
    \theta^{\tau - 5}\,
    c_{\tau}^{(2)}\,
    \LIR\LO_{\tau}^{J=2}(\n)+\ldots\,,
\end{align}
where the ellipsis includes terms with higher powers of $\Lambda_{\text{IR}}$ and $c^
{(2)}_{\tau}$ are again $O(1)$ OPE coefficients. Again $\tau$ spans the twist of the various light-ray operators entering in the OPE. For instance, in our theory and at leading order in $1/N$, double trace operators built out of two stress tensors have leading twist $\tau = 4$, while double trace operators built out of two scalar $\mathcal{O}$ have leading twist $\tau= 2\Delta_{\mathcal{O}}$ as in \eqref{Eq:DTOp}. 

When evaluated in a state $\ket{\mathcal{O}(p)}$, with $p^{\mu} = (p^0,\mathbf{0})$ in its rest frame for simplicity, we can readily estimate 
\begin{align}
    \braket{\En{\n_1}\En{\n_2}}\sim c^{(3)}_{\tau}\, \theta^{\tau-4}(p^0)^2 + c^{(2)}_{\tau}\,\LIR p^0 \theta^{\tau-5}\,,
\end{align}
where we are dropping $O(1)$ factors, comparing terms with the same $\tau$. We see that the IR truncation $\LIR$ becomes relatively more important as we go to small angles. 

A similar argument was proposed to estimate confinement effects in QCD \cite{Chen:2024nyc} but with less control of the conformal symmetry. The non-linearly realized conformal symmetry in our case allows us to systematically improve the small angle corrections as $\theta\rightarrow\LIR/p$.

The different methodology of Ref.~\cite{Csaki:2024joe} already finds a nontrivial angular dependence with a gap at leading order in $1/N$, in contradiction to our results. Indeed, we have shown that non-trivial angular dependence requires $1/N$ corrections leading to ``decays'', such as in our in-in loop process (see Fig.~\ref{fig:Loop} (a)), and we expect the gap to affect the resulting correlations at sufficiently small angles, as the hadronic gap does in real-world QCD (see, e.g., \cite{Chen:2024nyc}). 
However, such $1/N$ corrections were not included in Ref.~\cite{Csaki:2024joe}.  

\section{Outlook}\label{Sec:Outlook}

In this paper we studied energy correlator probes of 
strongly-coupled quantum field theory within their holographic dual description.
We set up these correlators in terms of in-in variants of Witten diagrams in 4D momentum
space and position space in the holographic direction, 
best suited to future phenomenological applications and physical collider intuition. We presented energy correlator computations in a manner that does not use conformal symmetry in any essential way so that we were able to   generalize away from exact CFT/AdS, 
deforming to non-trivial RG flow. 
We showed how scaling dimensions, dual to 5D particle masses in the holographic EFT description, can be predicted and measured from the angular dependence of energy correlators. 

We also generalized to the case of an IR 4D mass gap which is vital in any realistic setting, so that the BSM sector decays back to SM states before hitting detectors. 
We pointed out that the effects of the gap are not captured at fixed order in perturbation theory, even in the weakly coupled holographic description, but in RS1 they are parametrizable by a small angle OPE which exploits the fact that the gap corresponds to spontaneously broken conformal invariance in the dual CFT. It is likely that the OPE coefficients are actually calculable in RS1 EFT, but we did not attempt this here. We will tackle this challenge in future work, starting with the simpler problem of purely-4D weakly-coupled  models with spontaneous breaking of conformal invariance.

In realistic RS1 holographic models of BSM strong dynamics both types of deformations occur together. In particular, irrelevant or nearly marginal deformation away from conformal invariance are needed to couple to fields external to the strong dynamics and to trigger and stabilize the IR gap (such as famously accomplished by the Goldberger-Wise mechanism \cite{Goldberger:1999uk}) in RS1 and its CFT dual .
We plan study such a realistic setting in future work, in which we can use energy correlators to measure the non-trivial RG flow as a function of angular separation of detectors.  We wish to work towards a program of detective work  in which we can build up the significant features of the BSM strong dynamics from bottom up measurements. The holographic EFT formalism is well-suited to this incremental bottom-up exploration. We plan to study the power and limitations of AdS EFT in the energy correlator context in future work.\footnote{For example, in the classic strongly-coupled $N=4$ super-Yang-Mills theory explicit calculations of energy correlators from the AdS/CFT dual Supergravity EFT diverge for sufficiently small angles \cite{Chen:2024iuv}. Similarly in CFTs admitting a controlled large-charge expansion, where only low-dimension operators are retained (analogous to the implicit truncation of AdS EFT \cite{Fitzpatrick:2010zm}), the energy-correlator EFT calculation breaks down for sufficiently small angles \cite{Cuomo:2025pjp}.}
  We expect this to give insights into Lorentzian AdS/CFT duality. 

Even in asymptotically free BSM scenarios, energy correlators are inclusive enough to be insensitive (at large enough angles) to 
 the strong-coupling hadronization physics in the IR which may not be known in detail. 
We plan to study Hidden Valley \cite{Strassler:2006im} scenarios in which either type of BSM strong dynamics, whether approximately conformal or asymptotically free in the UV, can appear within upcoming High Luminosity LHC measurements. We will also explore hidden valley or UV extensions of the SM that may appear at future colliders. 

\vspace{1em}

\textit{\textbf{Acknowledgments}}
We would like to thank Z.~Chacko, G.~Cuomo, Yue-Zhou~Li, J.~Maldacena and 
I.~Moult for useful discussions. We also thank C.~Csaki and A.~Ismail for discussions related to their work. This work was supported by NSF grant PHY-2514660, NSF grant PHY-2210361 and by the Maryland Center for Fundamental Physics.

\bibliography{biblio} 

\newpage

\appendix
\onecolumngrid 

\end{document}